\documentclass[conference]{IEEEtran}
\IEEEoverridecommandlockouts
\usepackage{cite}
\usepackage{amsmath,amssymb,amsfonts}
\usepackage{algorithmic}
\usepackage{graphicx}
\usepackage{url}
\usepackage{textcomp}
\usepackage[table]{xcolor}
\def\BibTeX{{\rm B\kern-.05em{\sc i\kern-.025em b}\kern-.08em
    T\kern-.1667em\lower.7ex\hbox{E}\kern-.125emX}}
\usepackage{booktabs}
\usepackage{array}
\usepackage{makecell}
\usepackage{multirow}
\usepackage{placeins}
\usepackage[hidelinks]{hyperref}
\begin{document}

\title{TopoST: Topology-Conditioned Node Specialization for Urban Spatio-Temporal Foundation Models}

\author{
\IEEEauthorblockN{\makebox[0.30\textwidth][c]{1\textsuperscript{st} Kaiyun Wei}}
\IEEEauthorblockA{\textit{Faculty of Information Science} \\
\textit{and Computing} \\
\textit{University of Macau}\\
Macau, China \\
yc57929@um.edu.mo}
\and
\IEEEauthorblockN{\makebox[0.30\textwidth][c]{2\textsuperscript{nd} Yue Xu}}
\IEEEauthorblockA{\textit{Faculty of Information Science} \\
\textit{and Computing} \\
\textit{University of Macau}\\
Macau, China \\
yc57466@um.edu.mo}
\and
\IEEEauthorblockN{\makebox[0.30\textwidth][c]{3\textsuperscript{rd} Wenying Duan}}
\IEEEauthorblockA{\textit{School of Artificial} \\
\textit{Intelligence} \\
\textit{Nanchang University}\\
Nanchang, China \\
wenyingduan@ncu.edu.cn}
\and[\hfill\mbox{}\par\vspace{1em}\mbox{}\hfill]
\IEEEauthorblockN{\makebox[0.30\textwidth][c]{4\textsuperscript{th} Zichen Xu}}
\IEEEauthorblockA{\textit{School of Artificial} \\
\textit{Intelligence} \\
\textit{Nanchang University}\\
Nanchang, China \\
xuz@ncu.edu.cn‌}
\and
\IEEEauthorblockN{\makebox[0.30\textwidth][c]{5\textsuperscript{th} Shuai Ma}}
\IEEEauthorblockA{\textit{School of Computer Science} \\
\textit{and Engineering} \\
\textit{Beihang University}\\
Beijing, China\\
mashuai@buaa.edu.cn}
\and
\IEEEauthorblockN{\makebox[0.30\textwidth][c]{6\textsuperscript{th} Xiaoxi He}}
\IEEEauthorblockA{\textit{Faculty of Information Science} \\
\textit{and Computing} \\
\textit{University of Macau}\\
Macau, China \\
hexiaoxi@um.edu.mo}
}

\maketitle

\begin{abstract}
Urban spatio-temporal foundation models must share knowledge across diverse urban scenarios while accommodating heterogeneous node dynamics. Our analysis of 20+ datasets spanning diverse urban scenarios reveals recurring local graph structures across datasets alongside substantial node heterogeneity within local subgraphs. Motivated by these observations, we propose \textsc{TopoST}, which combines shared learning over local subgraphs with topology-conditioned node specialization. The key insight is to make node specialization itself transferable through a shared topology-to-parameter mapping. This mapping uses identity-independent topological representations to generate node-specific temporal and spatial dependency parameters, enabling direct specialization for unseen nodes without target-specific embeddings or model updates. Extensive experiments demonstrate that \textsc{TopoST} achieves remarkable zero-shot generalization across unseen cities and tasks, highlighting topology-conditioned node specialization as an effective mechanism for transferable urban spatio-temporal foundation models.
Code is available at \url{https://github.com/weikaiyun/TopoST}.
\end{abstract}


\section{Introduction}

Modern cities continuously generate spatio-temporal observations that capture complex urban dynamics, such as traffic flow and population movement. Accurate forecasting of these dynamics is essential for urban planning and operational decision-making. However, developing separate forecasting models for different urban scenarios requires repeated training and often substantial amounts of local data. Spatio-temporal foundation models (STFMs) offer a promising alternative by pretraining a shared predictor across multiple datasets and transferring the learned knowledge to new scenarios with limited or no parameter updates~\cite{liang2025foundation}. Yet achieving such transfer poses two intertwined challenges. Across datasets, spatial graphs differ in scale and topology, making graph-specific dependencies difficult to reuse. Within each graph, individual nodes exhibit distinct temporal dynamics and spatial interactions, making uniform dependency modeling insufficient. The fundamental challenge, therefore, is to develop a unified framework that enables effective knowledge transfer across graphs while accommodating node-specific dependencies within each graph.

Existing STFMs address spatio-temporal heterogeneity through different forms of knowledge sharing and adaptation. UniST~\cite{yuan2024unist} and UniFlow~\cite{yuan2026uniflow} retrieve reusable spatio-temporal patterns to construct prompts that adapt a shared backbone to local dynamics. OpenCity~\cite{li2025open} incorporates temporal context and spatial encodings into spatio-temporal dependency modeling, while UrbanFM~\cite{chen2026urbanfm} captures node-wise temporal dependencies and inter-node spatial interactions within local spatio-temporal units. FactoST~\cite{zhong2025factost,zhong2026learning} separates transferable temporal pretraining from spatial adaptation to target datasets. While these approaches provide different mechanisms for accommodating heterogeneous spatio-temporal data, we explore a novel perspective: \textbf{making node specialization itself transferable} across graphs. Specifically, can we learn a generalizable rule for constructing node-specific dependency models in unseen urban scenarios?

\begin{figure}[!t]
    \centering
    \includegraphics[width=\columnwidth]{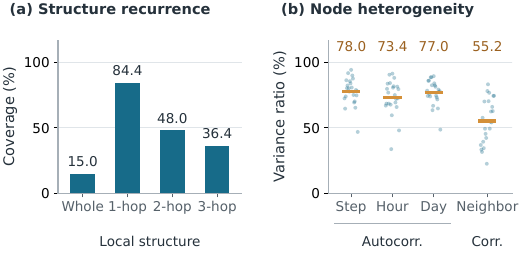}
    \caption{Structure recurrence and node heterogeneity across 20+ urban datasets (3,493 subgraphs; 55,737 nodes). Subgraphs are induced by METIS partitions~\cite{karypis1997metis} with a target size of 16 nodes per partition. (a) Coverage of recurring structures: the percentage of whole subgraphs or node-centered neighborhoods with an exact unweighted structural match in another dataset, averaged across datasets. (b) Within-subgraph variance ratios for temporal autocorrelation and mean neighbor correlation. Autocorrelation is evaluated at one-step, one-hour, and one-day lags. For each dataset, the ratio is defined as $R_{\mathrm{var}}=\frac{\sum_p\sum_{i\in\mathcal{V}_p}(a_i-\bar a_p)^2}{\sum_i(a_i-\bar a)^2}$, where $a_i$ is the statistic for node $i$, $\bar a_p$ is its mean within subgraph $\mathcal{V}_p$, and $\bar a$ is the overall mean. Dots denote datasets; orange marks and labels indicate the mean across datasets.}
    \label{fig:motivation}
\end{figure}

To investigate the feasibility of such transferable specialization, we analyze the structural and dynamic characteristics of 20+ datasets spanning diverse urban scenarios. Inspired by works on graph kernels~\cite{shervashidze2009graphlet, shervashidze2011weisfeiler}, which characterize graph similarity through local structural patterns, we examine the recurrence of local structures across datasets and the heterogeneity of node dynamics within local subgraphs. Our analysis reveals two observations that inform the design of a transferable specialization mechanism.

\textbf{Observation 1: Local graph structures recur across datasets, providing reusable contexts for shared learning.}
Although urban graphs differ substantially in their overall topologies, similar local connectivity patterns frequently emerge across datasets. As shown in Fig.~\ref{fig:motivation}(a), an average of 15.0\% of complete local subgraphs across datasets have an exact unweighted structural match in another dataset. At the neighborhood level, the recurrence rates are 84.4\%, 48.0\%, and 36.4\% when considering node-centered one-hop, two-hop, and three-hop neighborhoods, respectively. These results suggest that local structural patterns can be shared across datasets. Consequently, local subgraphs offer a natural unit for cross-dataset learning: a shared predictor can encounter recurring local connectivity patterns across otherwise distinct graphs and learn reusable dependency-modeling rules from their associated observations.

\textbf{Observation 2: Substantial node heterogeneity persists within local subgraphs, requiring node-specific dependency modeling.}
Structural recurrence across datasets does not imply uniform node dynamics within a local subgraph. To quantify their heterogeneity, we measure temporal autocorrelation at one-step, one-hour, and one-day lags, together with the mean correlation between each node and its neighbors. We then decompose the total node-level variation into within-subgraph and between-subgraph components. As illustrated in Fig.~\ref{fig:motivation}(b), averaging across datasets, within-subgraph differences account for 78.0\%, 73.4\%, and 77.0\% of the total variation in temporal autocorrelation across the three lags, respectively, and 55.2\% of the variation in mean neighbor correlation. These results reveal that nodes within the same local subgraph can exhibit markedly different temporal behaviors and spatial relationships. Therefore, sharing a predictor across local subgraphs should not require all constituent nodes to follow identical temporal and spatial dependency functions.

Together, these observations motivate a modeling strategy that shares dependency-modeling knowledge across local subgraphs, while specializing dependency parameters for individual nodes.
Training a shared predictor on local subgraphs enables it to learn from recurring local connectivity patterns across datasets, while node-specific dependency modeling preserves the flexibility needed to accommodate node heterogeneity within each context. Node-adaptive parameterization offers a potential mechanism for such specialization. For example, AGCRN~\cite{bai2020agcrn} generates node-specific model parameters from learned node embeddings and shared parameter pools. However, when these embeddings are indexed by training-node identities, they cannot directly parameterize unseen nodes in new graphs. This creates a critical gap between node-specific modeling and \emph{transferable} node-specific modeling: the latter requires a specialization mechanism that does not depend on previously encountered node identities.

We address this gap by using node-level topology as an identity-independent conditioning signal. Unlike learned identity embeddings, topological representations can be computed consistently for both pretrained and unseen graphs, capturing each node's local structural context. The central idea is thus to \textbf{transfer a topology-to-parameter mapping}, rather than transferring fixed node-specific parameters or fitting new node embeddings. This enables a pretrained model to construct specialized dependency functions for previously unseen nodes through a transferable rule.

\begin{figure*}[!t]
    \centering
    \includegraphics[width=0.95\textwidth]{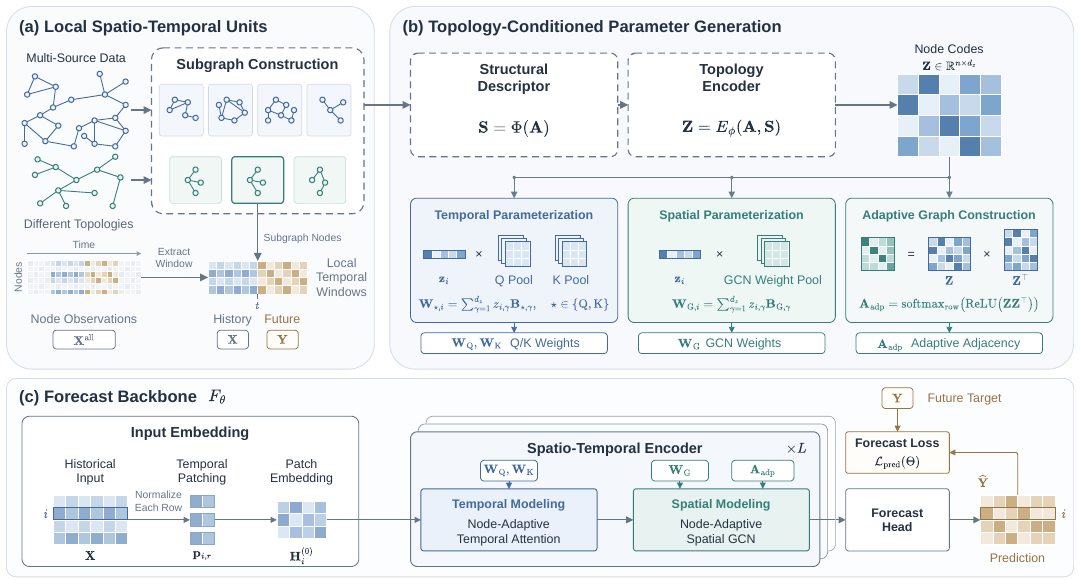}
    \caption{Overview of \textsc{TopoST}. (a) Spatial graphs are partitioned into local subgraphs, with observation rows for a selected subgraph highlighted in $\mathbf{X}^{\mathrm{all}}$. Windowing extracts historical inputs $\mathbf{X}$ ending at $t$ and subsequent future targets $\mathbf{Y}$ for supervision; each local input unit comprises a subgraph and its historical input $\mathbf{X}$. (b) Structural descriptors $\mathbf{S}=\Phi(\mathbf{A})$ and a shared topology encoder $E_{\phi}$ produce node codes $\mathbf{Z}$, which generate node-specific temporal query/key weights $\mathbf{W}_{\mathrm{Q}},\mathbf{W}_{\mathrm{K}}$ and spatial GCN weights $\mathbf{W}_{\mathrm{G}}$ from shared parameter pools, and construct an adaptive adjacency matrix $\mathbf{A}_{\mathrm{adp}}$. (c) The forecasting backbone $F_{\theta}$ embeds normalized historical inputs through temporal patching, linear projection, and sinusoidal positional encoding. A stack of $L$ spatio-temporal encoder blocks combines node-adaptive temporal attention with spatial graph convolution, followed by a forecast head producing $\widehat{\mathbf{Y}}$. All learnable components are jointly pretrained using the forecasting loss $\mathcal{L}_{\mathrm{pred}}(\Theta)$. Dataset, subgraph, forecast-origin, and block indices are omitted in this overview; each $\mathbf{W}_{\star}$ denotes the collection of node-specific weights $\{\mathbf{W}_{\star,i}\}_{i=1}^{n}$ for $\star\in\{\mathrm{Q},\mathrm{K},\mathrm{G}\}$.}
    \label{fig:framework}
\end{figure*}

Following this strategy, we propose \textsc{TopoST}, a spatio-temporal foundation model that combines training on local subgraphs with topology-conditioned node specialization (Fig.~\ref{fig:framework}). Specifically, we partition spatial graphs into local subgraphs and use their spatio-temporal observation windows to train a shared predictor across datasets. Within each subgraph, a shared topology encoder extracts node-level representations from local topology. These representations condition two separate parameter pools, which generate node-specific parameters for temporal and spatial dependency modeling, respectively. In this way, the predictor shares its overall architecture and parameter-generation rules across datasets while adapting its dependency computations to individual nodes. The topology encoder, parameter pools, and predictor are jointly pretrained on heterogeneous urban datasets. During inference, the same pretrained components directly generate dependency parameters for unseen nodes from their local topology, without learning target-specific node embeddings or updating model weights. TopoST therefore \textbf{achieves node-level specialization through transferable parameter generation} rather than target-specific adaptation.

Our main contributions are summarized as follows:
\begin{itemize}
    \item We make the novel observation that local connectivity patterns recur across datasets, while node dynamics remain substantially heterogeneous within local subgraphs. These observations are based on a comprehensive analysis of 20+ urban datasets and motivate a unified design that combines training on local subgraphs with explicit node-specific dependency modeling.

    \item We propose \textsc{TopoST}, a spatio-temporal foundation model that learns transferable topology-to-parameter mappings for node-specific dependency modeling. By combining training on local subgraphs with topology-conditioned parameter generation, TopoST enables transferable specialization for unseen nodes without relying on node identity embeddings or target-specific fine-tuning.

    \item Extensive experiments demonstrate that \textsc{TopoST} achieves remarkable zero-shot generalization across unseen cities and tasks. These results highlight topology-conditioned node specialization as an effective mechanism for transferring pretrained spatio-temporal models across heterogeneous urban scenarios.
\end{itemize}

\section{Related Work}

\subsection{Urban Spatio-Temporal Prediction}

Urban spatio-temporal prediction jointly models temporal dynamics and spatial interactions to forecast observations such as traffic and population flows. Deep forecasting models typically combine temporal sequence modeling with spatial dependency learning. For grid-based observations, ST-ResNet employs residual convolutional networks to capture spatial correlations together with temporal closeness, periodicity, and trends~\cite{zhang2017stresnet}. For graph-structured observations, DCRNN integrates diffusion graph convolution with recurrent encoder-decoder modeling~\cite{li2018dcrnn}, while STGCN combines graph convolution with gated temporal convolution~\cite{yu2018stgcn}. Graph WaveNet learns an adaptive adjacency matrix from node embeddings and couples graph convolution with dilated temporal convolution~\cite{wu2019graphwavenet}. Time-aware graph structure learning takes this idea further by allowing spatial dependencies to vary over time rather than relying on a fixed graph~\cite{ma2024timeaware}. Attention-based approaches such as STAEformer introduce adaptive spatio-temporal embeddings into Transformers to capture temporal and spatial dependencies~\cite{liu2023staeformer}. Spectral graph attention models such as STWave combine wavelet decomposition with efficient graph attention to disentangle temporal and spatial patterns~\cite{fang2023stwave}. Lightweight architectures such as LightCTS and BiST reduce model size and computation for correlated time series and spatio-temporal prediction~\cite{lai2023lightcts,ma2025bist}. These approaches provide effective forecasting architectures, with models typically trained separately for individual datasets. Reusing their predictive knowledge across heterogeneous urban scenarios motivates the development of urban spatio-temporal foundation models.

\subsection{Urban Spatio-Temporal Foundation Models}

Spatio-temporal foundation models learn reusable predictive knowledge by pretraining on heterogeneous datasets. UniST combines masked pretraining with knowledge-guided prompts to support prediction across urban scenarios~\cite{yuan2024unist}. UniFlow unifies grid-based and graph-based observations through multi-view patching and retrieves shared spatio-temporal patterns from memory~\cite{yuan2026uniflow}. UrbanDiT extends unified modeling to multiple tasks through diffusion Transformers with data-driven and task-specific prompts~\cite{yuan2025urbandit}. OpenCity combines Transformer and graph-based modeling with spatio-temporal context to support zero-shot traffic prediction~\cite{li2025open}. UrbanFM investigates scaling across data, computation, and architecture, organizing heterogeneous observations into local spatio-temporal units for shared dependency modeling~\cite{chen2026urbanfm}. FactoST and FactoST-v2 separate universal temporal pretraining from spatial adaptation to accommodate domain-specific relationships~\cite{zhong2025factost,zhong2026learning}. Scalable pretraining of compact urban spatio-temporal models further aims to reduce model size while exploiting large-scale multi-domain data~\cite{han2025scalable}. These studies provide different mechanisms for sharing knowledge and handling node-level heterogeneity across datasets. Within this pretraining paradigm, \textsc{TopoST} combines shared learning over local subgraphs with topology-conditioned parameter generation, enabling transferable specialization for unseen nodes without relying on node identity embeddings or target-specific fine-tuning.

\subsection{Node-Adaptive Modeling and Parameter Generation}

Conditional parameter generation accommodates heterogeneous node dynamics by allowing dependency functions to vary across locations. AGCRN combines learned node embeddings with shared parameter pools to generate node-specific graph-convolution parameters~\cite{bai2020agcrn}. ST-MetaNet extracts meta-knowledge from geographical and graph attributes to generate parameters for spatial attention and temporal recurrent networks~\cite{pan2019stmetanet}, while STWA generates location-specific and time-varying attention parameters from latent variables~\cite{cirstea2022stwa}. Parameter generation has also been explored for cross-city transfer: ST-GFSL derives non-shared parameters from node-level spatio-temporal meta-knowledge for few-shot learning~\cite{lu2022stgfsl}, and GPD pretrains a prompt-conditioned diffusion model on forecasting-network parameters learned from source cities to generate models for target scenarios~\cite{yuan2024gpd}. Building on this line of work, \textsc{TopoST} learns a topology-to-parameter mapping through joint pretraining on local subgraphs from heterogeneous datasets. Identity-independent topological representations condition separate temporal and spatial parameter pools, enabling the same pretrained rule to specialize dependency computations for unseen nodes without fitting target-specific embeddings or updating pretrained weights.

\section{Preliminaries}
\label{sec:preliminaries}


\subsection{Urban Spatio-Temporal Data}

Urban spatio-temporal data records observations that vary over time and across spatial locations, such as traffic flow, traffic speed, and mobility demand. We represent these observations as a tensor $\mathbf{X}^{\mathrm{all}}\in\mathbb{R}^{N\times T\times C}$, where $N$ is the number of spatial locations, $T$ the number of recorded time steps, and $C$ the number of measured features. Each entry $X^{\mathrm{all}}_{i,t,c}$ records feature $c$ at location $i$ and time $t$. Spatial locations may correspond to irregularly distributed sensors or cells in a regular grid.

We describe spatial relations using a graph $\mathcal{G}=(\mathcal{V},\mathcal{E},\mathbf{A})$, where $\mathcal{V}=\{1,\ldots,N\}$ is the set of locations, $\mathcal{E}$ is the edge set, and $\mathbf{A}\in\mathbb{R}_{\geq 0}^{N\times N}$ is the weighted adjacency matrix. The node ordering in $\mathbf{A}$ matches the spatial dimension of $\mathbf{X}^{\mathrm{all}}$.

\subsection{Urban Spatio-Temporal Task}

We study urban spatio-temporal forecasting across different urban scenarios and prediction horizons. Given historical observations $\mathbf{X}\in\mathbb{R}^{N\times T_h\times C}$ over $T_h$ time steps and their spatial adjacency $\mathbf{A}$, the task is to predict future observations $\mathbf{Y}\in\mathbb{R}^{N\times T_f\times C}$ over the next $T_f$ steps:
\begin{equation}
    \widehat{\mathbf{Y}}=\mathcal{F}_{\Theta}(\mathbf{X},\mathbf{A}),
    \label{eq:forecasting_problem}
\end{equation}
where $\mathcal{F}_{\Theta}$ denotes the complete forecasting model and $\Theta$ collects all its learnable parameters. Future measurements are unavailable as inputs.

\subsection{Urban Spatio-Temporal Foundation Models}

Urban spatio-temporal foundation models aim to learn reusable spatial and temporal knowledge from diverse urban datasets, supporting downstream tasks in new urban scenarios with limited or no task-specific supervision. Let $\{\mathcal{D}_m\}_{m=1}^{M}$ denote $M$ pretraining datasets spanning different regions and measured quantities. The goal is to learn a shared model $\mathcal{F}_{\Theta}$ that generalizes to an unseen dataset $\mathcal{D}'$ excluded from pretraining. The new dataset may differ in node count, graph topology, measured quantity, and temporal dynamics, without requiring correspondence between source and target nodes.

For historical observations $\mathbf{X}'$ and spatial adjacency $\mathbf{A}'$ from $\mathcal{D}'$, we consider two settings:

Zero-shot: Predict $\widehat{\mathbf{Y}}'=\mathcal{F}_{\Theta}(\mathbf{X}',\mathbf{A}')$ directly using the pretrained model, without dataset-specific training examples or parameter updates.

Few-shot: Fine-tune the pretrained model using a small set of labeled examples from $\mathcal{D}'$, obtaining adapted parameters $\Theta_{\mathrm{ft}}$, and predict $\widehat{\mathbf{Y}}'=\mathcal{F}_{\Theta_{\mathrm{ft}}}(\mathbf{X}',\mathbf{A}')$.

\section{Methodology}
\label{sec:methodology}

\subsection{Framework Overview}
\label{sec:framework_overview}

\textsc{TopoST} learns a shared topology-to-parameter mapping to make node specialization transferable. Figure~\ref{fig:framework} presents three stages: local spatio-temporal unit construction, topology-conditioned parameter generation, and shared spatio-temporal prediction. In \textsc{TopoST}, the complete model parameters are $\Theta=(\theta,\phi)$, where $\theta$ collects the forecasting backbone parameters, including the shared parameter pools, and $\phi$ contains the topology encoder parameters.

\textbf{Node-adaptive forecasting.}
Spatio-temporal forecasters can represent node-specific behavior through adaptation variables, including node embeddings, spatio-temporal identities, or coefficients for shared parameter pools~\cite{bai2020agcrn,liu2023staeformer}. We formulate node-adaptive forecasting as
\begin{equation}
    \widehat{\mathbf{Y}}
       =F_{\theta}(\mathbf{X},\mathbf{A};\mathbf{Z}),
    \label{eq:node_specific_interface}
\end{equation}
where $\theta$ contains the shared forecasting parameters and $\mathbf{Z}\in\mathbb{R}^{N\times d_z}$ contains node-specific variables. Each row $\mathbf{z}_i$ controls how the shared forecaster specializes its computation for node $i$. This formulation separates the common prediction mechanism from the variables that express differences among nodes.
When $\mathbf{Z}$ is learned as a free matrix indexed by node identity, its rows are tied to the training node set. An unseen target graph has no corresponding pretrained rows, even when its nodes occupy familiar structural contexts. Transferable node specialization therefore requires a shared rule for generating these variables from information available on both source and target graphs. Such a rule should allow the pretrained model to operate on a new node set without learning additional identity-indexed entries.

\textbf{Topology-conditioned transfer.}
\textsc{TopoST} achieves node-adaptive forecasting by generating node-specific temporal and spatial parameters from local topology. Local subgraphs provide contexts for learning across datasets, and a shared topology encoder generates the node-specific variables within each context. For a local subgraph $p$ containing $n_p$ nodes, let $\mathbf{X}_p$ denote its historical observations and $\mathbf{A}_p$ its adjacency matrix. Writing $\mathbf{S}_p=\Phi(\mathbf{A}_p)$ for its structural descriptors, we compute the node codes and forecasts as
\begin{equation}
\begin{aligned}
    &\mathbf{Z}_p=E_{\phi}(\mathbf{A}_p,\mathbf{S}_p),\\
    &\widehat{\mathbf{Y}}_p
       =\mathcal{F}_{\Theta}(\mathbf{X}_p,\mathbf{A}_p)\\
       &=F_{\theta}(\mathbf{X}_p,\mathbf{A}_p;\mathbf{Z}_p).
\end{aligned}
    \label{eq:framework_mapping}
\end{equation}
Here, $E_{\phi}$ maps the descriptors and local connectivity to node codes $\mathbf{Z}_p\in\mathbb{R}^{n_p\times d_z}$. Shared parameter pools use these codes to generate temporal query/key projections and spatial graph-convolution weights for each node. The codes also define an adaptive adjacency within the local group. Both parameter sets, $\theta$ and $\phi$, are jointly pretrained across source datasets.

On an unseen target graph, the same encoder computes its node codes, and the pretrained forecaster uses them with all learned parameters fixed. Local forecasts are assembled in the original node order to obtain predictions for the target dataset.

\subsection{Local Spatio-Temporal Units}
\label{sec:local_units}

We organize urban observations into local spatio-temporal units to support shared learning across datasets. Each unit pairs a local subgraph with a historical observation window over its nodes. Subsequent observations of the same nodes serve as forecasting targets. This formulation supports learning across graphs with heterogeneous scales and topologies without assuming correspondence between node identities or entire graphs.

To support shared dependency modeling and transferable node specialization, unit construction follows three design principles:
\begin{itemize}
    \item \textbf{Structural locality.} Node grouping should favor retaining graph connections within each unit while controlling the number of nodes it contains. This provides local contexts for shared dependency modeling across graphs of different sizes and topologies.
    \item \textbf{Topology preservation.} Within each selected node group, the unit should retain all original edges among its nodes, including their directions and weights. The resulting local graph supplies the structural information used to compute node codes for topology-conditioned parameter generation.
    \item \textbf{Construction consistency.} Unit construction should follow the same procedure across source and unseen target datasets, using the spatial structure and historical observations available in each dataset. The shared topology-to-parameter mapping therefore operates on consistently constructed local graphs during pretraining and inference.
\end{itemize}

Given an entire spatial graph $\mathcal{G}=(\mathcal{V},\mathcal{E},\mathbf{A})$ and its observation sequence $\mathbf{X}^{\mathrm{all}}$, the construction proceeds through spatial grouping, subgraph extraction, and temporal windowing, as illustrated in Fig.~\ref{fig:framework}(a).

\textbf{Step 1: Spatial grouping.}
We partition $\mathcal{V}$ into disjoint groups $\{\mathcal{V}_p\}_{p=1}^{P}$ satisfying $\bigcup_p\mathcal{V}_p=\mathcal{V}$. Spatial grouping can be based on geometric coordinates or graph connectivity. For example, PatchSTG~\cite{fang2025patchstg} uses a KDTree to group traffic sensors by geographic coordinates, while Cluster-GCN~\cite{chiang2019cluster} uses METIS to group graph nodes by connectivity for mini-batch training.

In \textsc{TopoST}, group boundaries determine which graph relations are available for structural encoding and node-specific parameter generation. We therefore seek disjoint, approximately balanced groups that retain local graph connections. We adopt METIS~\cite{karypis1997metis}, a multilevel graph partitioning method that seeks to minimize inter-partition edge cuts subject to approximate size balance. Reducing edge cuts favors retaining the relations used for topology conditioning, while size balance keeps local units comparable in node count. Its connectivity-based formulation also supports a common grouping procedure across source and target graphs.

METIS operates on the binary support $[\mathbf{A}_{\mathrm{bin}}]_{ij}=\mathbf{1}[A_{ij}>0\ \mathrm{or}\ A_{ji}>0]$, with self-loops removed. We set the number of partitions $P$ by dividing the total node count $N$ by the target group size and rounding up. Since METIS balances partitions approximately, the realized group sizes $n_p=|\mathcal{V}_p|$ may vary. Each group contains only its assigned nodes, without expansion to adjacent groups. Consequently, every node occurs in exactly one spatial group at a given forecast origin.

\textbf{Step 2: Subgraph extraction.}
The spatial context of group $\mathcal{V}_p$ is represented by its induced subgraph $\mathcal{G}_p=(\mathcal{V}_p,\mathcal{E}_p,\mathbf{A}_p)$:
\begin{equation}
\begin{aligned}
    &\mathcal{E}_p=\{(i,j)\in\mathcal{E}:i,j\in\mathcal{V}_p\},\\
    &\mathbf{A}_p=\mathbf{A}[\mathcal{V}_p,\mathcal{V}_p]
        \in\mathbb{R}^{n_p\times n_p}.
\end{aligned}
    \label{eq:local_graph}
\end{equation}
Extraction uses the original weighted adjacency, whose diagonal is zero. Hence, all within-group edges retain their directions and weights. We use $\mathbf{A}_p$ to compute structural descriptors and node codes for parameter generation. A normalized version of $\mathbf{A}_p$ supplies the fixed adjacency used in spatial graph convolution. The partitions and induced subgraphs are fixed preprocessing outputs and are reused across temporal windows.

\textbf{Step 3: Temporal windowing.}
For each group $\mathcal{V}_p$, we construct forecasting samples using sliding windows. At forecast origin $t$, we use the $T_h$ observations ending at $t$ as input and the next $T_f$ observations as forecasting targets for the same nodes. The local input unit and its targets are defined as
\begin{equation}
\begin{aligned}
    &\mathbf{X}_{p,t}=\mathbf{X}^{\mathrm{all}}[\mathcal{V}_p,t-T_h+1:t,:],\\
    &\mathbf{Y}_{p,t}=\mathbf{X}^{\mathrm{all}}[\mathcal{V}_p,t+1:t+T_f,:],\\
    &\mathcal{U}_{p,t}=(\mathcal{V}_p,\mathbf{A}_p,\mathbf{X}_{p,t}).
\end{aligned}
    \label{eq:local_st_unit}
\end{equation}
The time intervals include both endpoints, so that $\mathbf{X}_{p,t}\in\mathbb{R}^{n_p\times T_h\times C}$ and $\mathbf{Y}_{p,t}\in\mathbb{R}^{n_p\times T_f\times C}$. We use the same node ordering in $\mathbf{A}_p$, $\mathbf{X}_{p,t}$, and $\mathbf{Y}_{p,t}$. When temporal splits are used, input--target windows are constructed entirely within each split. Only $\mathcal{U}_{p,t}$ is supplied as the local forecasting input; $\mathbf{Y}_{p,t}$ is used for supervision. The input construction therefore remains applicable at inference without access to future measurements.

\textbf{Training and inference.}
We construct $\mathcal{D}_{\mathrm{train}}$ by pooling $(\mathcal{U}_{p,t},\mathbf{Y}_{p,t})$ extracted from the source sequences designated for pretraining. Each mini-batch comprises temporal windows from a single induced subgraph, accommodating variable group sizes without spatial padding. The same construction is applied to unseen target graphs. Because the groups form a disjoint cover, their local forecasts can be assembled into a dataset-level prediction in the original node order.

\subsection{Structural Node Descriptors}
\label{sec:structural_conditioning}
\label{sec:local_descriptor}

We train a shared forecasting model on local spatio-temporal units from multiple datasets. To specialize this shared model to individual nodes, we require conditioning information that captures local structural variation and remains computable on unseen graphs. The recurrence of local connectivity patterns in Fig.~\ref{fig:motivation}(a) motivates using structural measurements as a common basis for specialization. We can use local structural encodings (Local SEs) to characterize each node through neighborhood statistics, random-walk, heat-kernel diffusion, and local curvature~\cite{rampasek2022gps}.

For structural encoding and forecasting below, we describe a single local unit: we omit its dataset, subgraph, and forecast-origin indices and write $n=n_p$, $\mathbf{A}=\mathbf{A}_p$, $\mathbf{X}=\mathbf{X}_{p,t}$, and $\mathbf{Y}=\mathbf{Y}_{p,t}$. Node indices $i,j$ refer to the local ordering. We compute
\begin{equation}
    \mathbf{S}=\Phi(\mathbf{A})\in\mathbb{R}^{n\times d_s},
    \label{eq:structural_descriptor}
\end{equation}
where row $\mathbf{s}_i$ describes node $i$, and $\Phi$ is deterministic and nontrainable. We apply the same encoding operator $\Phi$ to all source and target graphs, yielding node descriptors with a fixed dimension $d_s$ independent of graph size. Descriptors for unseen nodes are computed directly from their graph, without source--target node correspondence or newly fitted variables.

To ensure that these structural inputs do not depend on arbitrary node labels, we require permutation equivariance:
\begin{equation}
    \Phi(\boldsymbol{\Pi}\mathbf{A}\boldsymbol{\Pi}^{\top})=\boldsymbol{\Pi}\Phi(\mathbf{A}),
    \label{eq:descriptor_equivariance}
\end{equation}
for any permutation matrix $\boldsymbol{\Pi}$. Relabeling only reorders the rows of $\mathbf{S}$, preserving each node's descriptor.

We consider four local structural encodings as alternative choices for $\Phi$, each characterizing nodes through a different set of structural measurements.

Let $\mathbf{A}_{\mathrm{sym}}=(\mathbf{A}+\mathbf{A}^{\top})/2$ denote the symmetrized adjacency matrix, $\mathcal{N}(i)=\{j:[\mathbf{A}_{\mathrm{sym}}]_{ij}>0\}$ the neighborhood of node $i$, and $\deg(i)=|\mathcal{N}(i)|$ its unweighted degree. The corresponding weighted degree matrix $\mathbf{D}_{\mathrm{sym}}$ is diagonal, with $[\mathbf{D}_{\mathrm{sym}}]_{ii}=\sum_{j=1}^{n}[\mathbf{A}_{\mathrm{sym}}]_{ij}$. Inverse degree entries for isolated nodes are set to zero.

\textbf{Local structural encodings.}
We consider four structural encoders, each producing a node descriptor $\mathbf{s}_i\in\mathbb{R}^{d_s}$.
\textbf{SSE} (Statistical Structural Encoding) collects $d_s$ local graph statistics, such as node degree, neighboring-degree summaries and the local clustering coefficient~\cite{rampasek2022gps}.
\textbf{RWSE} (Random-Walk Structural Encoding) records return probabilities as $\mathbf{s}_i^{\mathrm{RWSE}}=[(\mathbf{P}_{\mathrm{rw}}^k)_{ii}]_{k=1}^{d_s}$, where $\mathbf{P}_{\mathrm{rw}}=\mathbf{D}_{\mathrm{sym}}^{-1}\mathbf{A}_{\mathrm{sym}}$~\cite{dwivedi2022lspe}.
\textbf{HKSE} (Heat-Kernel Structural Encoding) measures heat retention at prescribed diffusion times using $\mathbf{s}_i^{\mathrm{HKSE}}=[(\exp(-\tau_k\mathbf{L}))_{ii}]_{k=1}^{d_s}$, where $\mathbf{L}=\mathbf{D}_{\mathrm{sym}}-\mathbf{A}_{\mathrm{sym}}$~\cite{sun2009hks}.
\textbf{FRSE} (Forman-Ricci Structural Encoding) applies $d_s$ permutation-invariant summary statistics, such as the mean, standard deviation, minimum and maximum to the multiset of incident-edge curvatures on the unweighted support graph~\cite{topping2022understanding}.
For each structural encoding, the statistics, walk lengths or diffusion times defining its coordinates are fixed across source and target graphs.

We use SSE as the default and compare all four descriptor choices in Section~\ref{sec:ablation_studies}. The SSE descriptors are computed once per induced subgraph and supplied as input node features to the shared topology encoder.

\subsection{Topology Encoder}
\label{sec:topology_encoder}

Given the structural descriptors $\mathbf{S}$, the topology encoder $E_{\phi}$ generates node codes that condition temporal and spatial parameter generation:
\begin{equation}
    \mathbf{Z}=E_{\phi}(\mathbf{A},\mathbf{S})
       \in\mathbb{R}^{n\times d_z}.
    \label{eq:topology_codes}
\end{equation}
The encoder maps each node's structural context to its specialization variables. Joint pretraining with the forecasting backbone learns how these structural inputs should inform node-specific dependency modeling.

\textbf{Encoder design principles.}
Cross-dataset parameter generation calls for node codes with three complementary properties: permutation equivariance maintains node correspondence under relabeling; structural discriminability preserves relevant differences between local structural contexts; and inductive generalization allows the same pretrained encoder to generate codes for unseen nodes across graphs of different sizes.

\textbf{Permutation equivariance.}
The generated codes should preserve node correspondence under relabeling. For any permutation matrix $\boldsymbol{\Pi}$, we require
\begin{equation}
    E_{\phi}(\boldsymbol{\Pi}\mathbf{A}\boldsymbol{\Pi}^{\top},\boldsymbol{\Pi}\mathbf{S})
       =\boldsymbol{\Pi}E_{\phi}(\mathbf{A},\mathbf{S}).
    \label{eq:encoder_equivariance}
\end{equation}
Together with descriptor equivariance in~\eqref{eq:descriptor_equivariance}, this ensures that relabeling the graph only reorders its node codes. This symmetry requirement establishes consistency under node relabeling; the next principle concerns the structural distinctions retained by the codes.

\textbf{Structural discriminability.}
We use one-dimensional Weisfeiler--Lehman (1-WL) refinement as a reference for comparing node-centered structural contexts~\cite{shervashidze2011weisfeiler}. At each iteration, 1-WL updates a node's color using its current color and the multiset of neighboring colors. A continuous analogue starts from $\mathbf{u}_i^{(0)}=\mathbf{s}_i$ and refines node representations through shared maps $f_{k}(\mathbf{u},\mathcal{M})$, where $\mathbf{u}$ is the current node representation and $\mathcal{M}$ is the multiset of neighboring representations. Sharing these maps allows recurring structural contexts to be processed by the same refinement rule across datasets. Injectivity provides a sufficient condition for retaining distinctions between different input pairs~\cite{xu2019gin}:
\begin{equation}
\begin{aligned}
    &(\mathbf{u},\mathcal{M})\ne(\mathbf{u}',\mathcal{M}')\\
    &\Longrightarrow\quad
       f_{k}(\mathbf{u},\mathcal{M})
       \ne f_{k}(\mathbf{u}',\mathcal{M}').
\end{aligned}
    \label{eq:topology_refinement}
\end{equation}
We use this criterion as a design reference for informative neighborhood refinement.

\textbf{Inductive generalization.}
The same refinement functions and output mapping are shared across all nodes and local graphs, making the structural rule applicable beyond the training node set~\cite{hamilton2017graphsage}. After $K$ refinement layers, a shared node-wise readout produces the conditioning codes:
\begin{equation}
    \mathbf{z}_i=r_{\phi}\!\left(\mathbf{u}_i^{(K)}\right)
       \in\mathbb{R}^{d_z},
    \label{eq:topology_readout}
\end{equation}
where $r_{\phi}:\mathbb{R}^{d_K}\to\mathbb{R}^{d_z}$ maps the final node representation to the conditioning dimension. All parameter dimensions are independent of node count. The pretrained encoder therefore computes codes for unseen target nodes directly from their local graphs, with $\phi$ fixed.

\textbf{Instantiation.}
We instantiate $E_{\phi}$ with the graph isomorphism network (GIN), which can match the node-level discriminative power of 1-WL under injective neighborhood aggregation and updates~\cite{xu2019gin}. This connects the encoder choice to the refinement criterion above. Let $\mathbf{u}_i^{(k)}\in\mathbb{R}^{d_k}$ denote the representation of node $i$ at layer $k$, initialized by $\mathbf{u}_i^{(0)}=\mathbf{s}_i$ with $d_0=d_s$. The neighborhood $\mathcal{N}(i)$ is defined by $\mathbf{A}_{\mathrm{sym}}$ as above. At layer $k=0,\ldots,K-1$, GIN updates each node representation as
\begin{equation}
    \mathbf{u}_i^{(k+1)}
       =\operatorname{MLP}_{k}\!\left(
          (1+\epsilon_k)\mathbf{u}_i^{(k)}
          +\sum_{j\in\mathcal{N}(i)}\mathbf{u}_j^{(k)}
          \right).
    \label{eq:topology_gin}
\end{equation}
Here, $\epsilon_k$ is a fixed or learnable scalar controlling the contribution of the central node. Each multilayer perceptron $\operatorname{MLP}_{k}$ maps $\mathbb{R}^{d_k}$ to $\mathbb{R}^{d_{k+1}}$ and is shared across nodes and graphs. The final representations are mapped to node codes by~\eqref{eq:topology_readout}. These codes condition the shared parameter pools introduced next, linking each node's structural context to its temporal and spatial forecasting operators.

\subsection{Topology-Conditioned Dependency Modeling}
\label{sec:parameter_generation}
\label{sec:dependency_modeling}

The structural recurrence in Fig.~\ref{fig:motivation}(a) motivates sharing dependency-modeling knowledge across local subgraphs, while the within-subgraph heterogeneity in Fig.~\ref{fig:motivation}(b) calls for node-specific temporal and spatial dependency functions. We reconcile these requirements through a topology-to-parameter mapping: topology-derived node codes condition the generation of dependency parameters, making the specialization rule applicable to unseen nodes.

The forecasting backbone first applies \textbf{instance normalization} to each node's historical observations and divides the normalized sequence into temporal patches. A linear projection and sinusoidal positional encoding yield the patch embeddings $\mathbf{H}_i^{(0)}\in\mathbb{R}^{T_{\mathrm{p}}\times d}$, where $T_{\mathrm{p}}$ is the number of temporal patches and $d$ is the embedding dimension. These embeddings pass through $L$ stacked spatio-temporal blocks, each comprising temporal self-attention, spatial graph convolution, and a feed-forward network. Each block uses the same node codes to generate temporal and spatial parameters from its own parameter pools.

\textbf{Topology-conditioned parameter generation.}
We express node-specific dependency operators as linear combinations of basis matrices~\cite{bai2020agcrn}. For block $\ell=1,\ldots,L$, node $i$ uses its topology code to generate the query, key, and graph-convolution projections:
\begin{equation}
    \mathbf{W}_{\star,i}^{(\ell)}
      =\sum_{\gamma=1}^{d_z}z_{i,\gamma}
        \mathbf{B}_{\star,\gamma}^{(\ell)},
    \qquad \star\in\{\mathrm{Q},\mathrm{K},\mathrm{G}\}.
    \label{eq:conditioned_parameters}
\end{equation}
Here, $\mathbf{B}_{\star,\gamma}^{(\ell)}\in\mathbb{R}^{d\times d}$ are trainable basis matrices, with an independent pool for each projection. Each pool is shared across nodes and datasets, and the coefficients $z_{i,\gamma}$ are computed from each node's structural context. Query and key biases are generated analogously from vector pools; the graph projection has no bias. The generated parameters apply to all patches of a node. For clarity, we describe the operations within a single block below and omit its index.

\textbf{Node-adaptive temporal attention.}
The observed heterogeneity in temporal autocorrelation motivates allowing nodes to differ in how they relate historical patches. In temporal self-attention, query--key dot products determine the attention weights assigned to these patches. We therefore generate node-specific query and key projections to specialize the temporal matching function according to structural context. Given the block input $\mathbf{H}_i\in\mathbb{R}^{T_{\mathrm{p}}\times d}$, we compute
\begin{equation}
\begin{aligned}
    &\overline{\mathbf{H}}_i=\operatorname{RMSNorm}(\mathbf{H}_i),\\
    &\mathbf{Q}_i=\overline{\mathbf{H}}_i\mathbf{W}_{\mathrm{Q},i},\quad
    \mathbf{K}_i=\overline{\mathbf{H}}_i\mathbf{W}_{\mathrm{K},i},\\
    &\mathbf{V}_i=\overline{\mathbf{H}}_i\mathbf{W}_{\mathrm{V}},\\
    &\mathbf{H}_i^{\mathrm{T}}=\operatorname{RMSNorm}\!\left(
      \overline{\mathbf{H}}_i+
      \operatorname{MHA}(\mathbf{Q}_i,\mathbf{K}_i,\mathbf{V}_i)\right).
\end{aligned}
    \label{eq:temporal_attention}
\end{equation}
Here, $\overline{\mathbf{H}}_i$ denotes the normalized patch embeddings, and $\mathbf{W}_{\mathrm{V}}$ is shared across nodes and datasets. We use RMSNorm throughout the forecasting blocks. Bias terms are omitted from the displayed projections, and $\operatorname{MHA}$ concatenates the outputs of scaled dot-product attention heads over temporal patches. Topology thus conditions how temporal relevance is evaluated, and the resulting attention weights depend jointly on this structural context and the observed sequence.

\textbf{Node-adaptive spatial graph convolution.}
The heterogeneity in neighbor correlations similarly motivates node-specific spatial dependency modeling. We specialize both how nodes aggregate information and how they transform the aggregated messages. For aggregation, we combine observed connectivity with additional propagation weights derived from the node codes:
\begin{equation}
\begin{aligned}
    &\mathbf{A}_{\mathrm{fix}}=\widetilde{\mathbf{D}}^{-1/2}
      (\mathbf{A}+\mathbf{I})\widetilde{\mathbf{D}}^{-1/2},\\
    &\mathbf{A}_{\mathrm{adp}}=\operatorname{Softmax}
      \!\left(\operatorname{ReLU}(\mathbf{Z}\mathbf{Z}^{\top})\right),\\
    &\mathbf{A}_{\mathrm{mix}}=\mathbf{A}_{\mathrm{fix}}
      +\lambda\mathbf{A}_{\mathrm{adp}},
\end{aligned}
    \label{eq:spatial_graphs}
\end{equation}
Here, $\mathbf{I}$ is the identity matrix, $\widetilde{\mathbf{D}}$ is diagonal with $\widetilde{D}_{ii}=\sum_j[\mathbf{A}+\mathbf{I}]_{ij}$, and $\operatorname{Softmax}$ normalizes each row over sending nodes. The fusion scalar $\lambda$ is learned separately in each block. The fixed graph preserves observed relations, and the adaptive term uses similarities between structural codes to weight additional interactions within the local unit. To let receiving nodes respond differently to aggregated information, we apply the generated projection $\mathbf{W}_{\mathrm{G},i}$ after aggregation:
\begin{equation}
\begin{aligned}
    &\mathbf{G}_i=\left(\sum_{j=1}^{n}
      [\mathbf{A}_{\mathrm{mix}}]_{ij}\mathbf{H}_j^{\mathrm{T}}\right)
      \mathbf{W}_{\mathrm{G},i},\\
    &\mathbf{H}_i^{\mathrm{S}}=\mathbf{H}_i
      +\operatorname{Linear}\!\left(
        \alpha\mathbf{H}_i^{\mathrm{T}}+(1-\alpha)\mathbf{G}_i\right).
\end{aligned}
    \label{eq:spatial_convolution}
\end{equation}
Here, $\mathbf{G}_i\in\mathbb{R}^{T_{\mathrm{p}}\times d}$ contains the transformed messages for all patches of node $i$, $\alpha$ is a mixing coefficient, and $\operatorname{Linear}$ applies a linear projection to each patch embedding. The residual connection adds the original block input $\mathbf{H}_i$ to obtain the spatial output $\mathbf{H}_i^{\mathrm{S}}$. Together, $\mathbf{A}_{\mathrm{mix}}$ and $\mathbf{W}_{\mathrm{G},i}$ condition spatial aggregation and message transformation on structural context, supporting node-specific spatial computations.

\textbf{Block output.}
A feed-forward network produces the block output through a normalized residual update:
\begin{equation}
    \mathbf{H}_i^{\mathrm{out}}=\mathbf{H}_i^{\mathrm{S}}
       +\operatorname{FFN}\!\left(
          \operatorname{RMSNorm}\!\left(
             \mathbf{H}_i^{\mathrm{S}}\right)\right).
    \label{eq:block_output}
\end{equation}
Normalization and the feed-forward network operate on each patch embedding. The block output $\mathbf{H}_i^{\mathrm{out}}$ becomes the input to the next block.

\textbf{Prediction.}
After $L$ blocks, we flatten each node's final patch embeddings $\mathbf{H}_i^{(L)}$ into a row of $\mathbf{H}_{\mathrm{flat}}^{(L)}\in\mathbb{R}^{n\times(T_{\mathrm{p}}d)}$. A linear prediction head maps these representations to forecasts for all nodes. We retain the first $T_f$ output steps and restore the original measurement scale:
\begin{equation}
    \widehat{\mathbf{Y}}
      =\boldsymbol{\mu}^{\top}+\boldsymbol{\sigma}^{\top}\odot
        \operatorname{Head}\!\left(
          \mathbf{H}_{\mathrm{flat}}^{(L)}\right)_{:,1:T_f}.
    \label{eq:forecast_head}
\end{equation}
Here, $\widehat{\mathbf{Y}}\in\mathbb{R}^{n\times T_f}$ is the prediction matrix for the single-channel setting. The row vectors $\boldsymbol{\mu},\boldsymbol{\sigma}\in\mathbb{R}^{1\times n}$ collect the mean and standard deviation of each node's historical observations. The operator $\odot$ denotes element-wise multiplication, with the normalization statistics broadcast across the forecast horizon.

\textbf{Joint pretraining.}
We jointly optimize the topology encoder, parameter pools, and forecasting backbone using local historical--future pairs from the source datasets. Each dataset provides one measurement channel ($C=1$). For a mini-batch $\mathcal{B}$ sampled from $\mathcal{D}_{\mathrm{train}}$, let $\widehat{\mathbf{Y}}_{\mathcal{B}}$ and $\mathbf{Y}_{\mathcal{B}}$ denote the batch forecasts and targets, respectively. We minimize the mean absolute error (MAE):
\begin{equation}
    \mathcal{L}_{\mathrm{pred}}(\Theta)
      =\mathbb{E}_{\mathcal{B}\sim\mathcal{D}_{\mathrm{train}}}\!\left[
        \operatorname{MAE}\!\left(
          \widehat{\mathbf{Y}}_{\mathcal{B}},\mathbf{Y}_{\mathcal{B}}
        \right)\right].
    \label{eq:training_objective}
\end{equation}
MAE averages the absolute prediction errors over the future measurements retained for supervision, using their original measurement scale. Through the generated operators, prediction errors train the topology-to-parameter mapping to condition temporal matching and spatial message transformation on structural context. Partitioning, subgraph extraction, and structural descriptor computation remain fixed. On an unseen graph, the pretrained topology encoder and parameter pools directly construct dependency operators from its local topology, without learning target-specific node embeddings or updating model weights. Thus, node specialization itself becomes transferable across datasets.

\begin{table*}[!t]
    \centering
    \caption{Forecasting results over a \textbf{one-day horizon} on the OpenCity benchmark. UrbanFM and UniST use their official implementations. OpenCity and full-shot baseline results are taken from the original OpenCity paper. Lower values are better. Bold and underlined values indicate the best and second-best results in each column, respectively. MAPE is reported in percent.}
    \label{tab:zero-shot}
    \begingroup
    \footnotesize
    \setlength{\tabcolsep}{2pt}
    \renewcommand{\arraystretch}{1.0}
    \resizebox{\textwidth}{!}{%
    \begin{tabular}{@{}ll*{18}{w{c}{3.6em}}@{}}
        \toprule
        \multirow{2}{*}{Setting} & \multirow{2}{*}{Model} & \multicolumn{3}{c}{CAD3} & \multicolumn{3}{c}{CAD5} & \multicolumn{3}{c}{PEMS07M} & \multicolumn{3}{c}{TrafficSH} & \multicolumn{3}{c}{CHI-TAXI} & \multicolumn{3}{c}{NYC-BIKE} \\
        \cmidrule(lr){3-5}\cmidrule(lr){6-8}\cmidrule(lr){9-11}\cmidrule(lr){12-14}\cmidrule(lr){15-17}\cmidrule(lr){18-20}
         & & MAE & RMSE & MAPE & MAE & RMSE & MAPE & MAE & RMSE & MAPE & MAE & RMSE & MAPE & MAE & RMSE & MAPE & MAE & RMSE & MAPE \\
        \midrule
        \multirow{10}{*}{Full-shot} & PDFormer & 20.28 & 36.43 & 25.19 & 12.89 & 21.18 & 28.82 & 4.62 & 8.36 & 13.74 & 0.77 & 1.23 & 8.34 & 4.03 & 12.82 & 44.42 & 7.33 & 13.01 & 65.44 \\
          & STWA & 21.65 & 37.55 & 26.85 & 14.43 & 24.14 & 29.34 & 4.54 & 8.57 & 12.91 & 1.42 & 2.49 & 9.92 & 3.70 & 11.49 & 42.52 & 7.97 & 14.35 & 64.19 \\
          & MSDR & 22.74 & 37.15 & 32.37 & 15.44 & 29.52 & 32.49 & 4.81 & 8.61 & 13.90 & 1.84 & 3.55 & 9.91 & 3.55 & 10.39 & 52.88 & 8.08 & 13.97 & 77.56 \\
          & AGCRN & 18.72 & 31.93 & 25.45 & 12.88 & 23.78 & 26.29 & 4.61 & 8.63 & 12.81 & 1.33 & 2.45 & 8.51 & 3.60 & 11.31 & 46.48 & 8.13 & 14.46 & 78.36 \\
          & MTGNN & 17.59 & 28.92 & 25.22 & 11.70 & 20.30 & 26.75 & 4.52 & 8.06 & 13.11 & 0.81 & 1.26 & 8.29 & 3.27 & 9.87 & \underline{39.38} & 6.48 & 11.49 & \underline{61.52} \\
          & STSGCN & 21.88 & 34.52 & 30.20 & 13.87 & 22.32 & 32.84 & 4.56 & 8.05 & 12.70 & 1.66 & 3.33 & 9.33 & 4.87 & 14.40 & 104.64 & 6.85 & 11.98 & 68.44 \\
          & GWN & 16.94 & 28.81 & 22.98 & 10.69 & 19.75 & \underline{25.98} & \underline{4.17} & \underline{7.84} & \underline{11.46} & 0.76 & 1.39 & 9.23 & 3.56 & 11.27 & 41.31 & 7.61 & 13.56 & \textbf{58.41} \\
          & ASTGNN & 23.60 & 39.35 & 41.22 & 12.58 & 21.23 & 30.56 & 4.39 & 8.21 & 12.58 & 0.69 & 1.09 & 8.05 & 3.28 & 10.32 & 42.82 & 6.44 & 11.62 & 63.90 \\
          & STGCN & 20.24 & 34.34 & 25.33 & 13.76 & 27.49 & 32.89 & 4.44 & 8.37 & 12.59 & 1.60 & 3.14 & 8.04 & 3.09 & 9.54 & 42.47 & 8.01 & 13.94 & 63.02 \\
          & TGCN & 19.56 & 30.82 & 28.15 & 13.07 & 21.56 & 32.65 & 4.88 & 8.38 & 13.94 & 1.79 & 2.65 & 17.75 & 4.02 & 11.70 & 60.25 & 7.75 & 13.49 & 85.27 \\
        \midrule
        \multirow{4}{*}{Zero-shot} & UniST & 24.22 & 42.80 & 30.39 & 14.78 & 25.61 & 34.87 & 4.72 & 9.03 & 11.75 & 0.63 & 1.13 & 8.74 & 2.75 & 6.02 & 72.26 & 6.93 & 12.25 & 77.01 \\
          & OpenCity & 15.88 & 27.03 & 21.94 & 11.09 & 18.96 & 27.90 & 4.50 & 8.21 & 12.20 & \underline{0.55} & \underline{0.85} & \underline{8.01} & \underline{1.91} & \underline{4.42} & 40.07 & 6.32 & 11.60 & 70.06 \\
          & UrbanFM & \underline{14.74} & \underline{24.66} & \underline{21.43} & \underline{10.53} & \underline{18.10} & 26.71 & 4.45 & 8.25 & 12.52 & 0.56 & 0.86 & 8.16 & 1.96 & 4.44 & 42.65 & \underline{6.14} & \underline{11.03} & 68.21 \\
          & \textbf{TopoST} & \textbf{13.61} & \textbf{21.98} & \textbf{20.29} & \textbf{9.36} & \textbf{15.82} & \textbf{24.12} & \textbf{4.11} & \textbf{7.41} & \textbf{10.48} & \textbf{0.51} & \textbf{0.79} & \textbf{7.41} & \textbf{1.84} & \textbf{4.19} & \textbf{38.92} & \textbf{5.35} & \textbf{9.57} & 63.46 \\
        \bottomrule
    \end{tabular}%
    }
    \endgroup
\end{table*}

\begin{table*}[!t]
    \centering
    \caption{Zero-shot forecasting on the 12 target datasets of the UrbanFM benchmark. Each cell reports results for \textbf{12$\to$12} / \textbf{24$\to$24} input--output steps. UrbanFM, FactoST, and OpenCity results are taken from the original UrbanFM paper. UniST uses its official implementation. Lower is better; bold and underlined values indicate the best and second-best results for each dataset, metric, and horizon, respectively. MAPE is in percent.}
    \label{tab:urbanfm-zero-shot}
    \begingroup
    \footnotesize
    \setlength{\tabcolsep}{1.5pt}
    \renewcommand{\arraystretch}{1.08}
    \newlength{\urbanfmcolumnwidth}
    \settowidth{\urbanfmcolumnwidth}{\,/\,}
    \addtolength{\urbanfmcolumnwidth}{5.6em}
    \resizebox{\textwidth}{!}{%
    \begin{tabular}{@{}*{2}{w{c}{\urbanfmcolumnwidth}}*{12}{w{r}{2.8em}@{\,/\,}w{l}{2.8em}}@{}}
        \toprule
        Model & Metric $\downarrow$ & \multicolumn{2}{c}{PEMS-03} & \multicolumn{2}{c}{PEMS-04} & \multicolumn{2}{c}{PEMS-07} & \multicolumn{2}{c}{PEMS-08} & \multicolumn{2}{c}{\makecell{OCC-\\Paris}} & \multicolumn{2}{c}{\makecell{OCC-\\Hamburg}} & \multicolumn{2}{c}{\makecell{PEMS-\\BAY}} & \multicolumn{2}{c}{METR-LA} & \multicolumn{2}{c}{\makecell{Traffic-\\SH}} & \multicolumn{2}{c}{\makecell{Bike-\\NYC}} & \multicolumn{2}{c}{\makecell{Taxi-\\NYC}} & \multicolumn{2}{c}{\makecell{TDrive-\\BJ}} \\
        \midrule
        \multirow{3}{*}{UniST} & MAE & 25.22 & 35.39 & 34.63 & 46.26 & 44.02 & 56.86 & 29.88 & 39.89 & 1.79 & 2.21 & 6.66 & 7.41 & 2.18 & 3.18 & 5.93 & 8.27 & 0.60 & 0.73 & 3.33 & 4.81 & \underline{5.21} & \underline{6.02} & 37.93 & 40.48 \\
         & RMSE & 38.10 & 55.04 & 50.69 & 68.29 & 62.03 & 82.41 & 42.93 & 57.78 & 3.16 & 4.01 & 10.98 & 12.12 & 4.76 & 6.68 & 12.65 & 15.75 & 0.84 & 1.05 & 6.54 & 9.99 & 14.36 & 17.24 & 72.97 & 77.80 \\
         & MAPE & 21.30 & 28.42 & 22.49 & 28.99 & 19.08 & 23.83 & 18.28 & 23.42 & 25.59 & 30.51 & 55.63 & 62.52 & 4.26 & 6.43 & 21.14 & 30.44 & 7.48 & 9.21 & 42.11 & 54.29 & 30.63 & \underline{34.44} & 51.43 & 54.04 \\
        \midrule
        \multirow{3}{*}{OpenCity} & MAE & 25.57 & 27.85 & 32.82 & 35.36 & 34.50 & 38.41 & 26.92 & 29.64 & 1.80 & 2.03 & 7.07 & 7.33 & 2.22 & 2.68 & 5.44 & 6.38 & 0.45 & \underline{0.48} & 4.67 & 5.17 & 12.23 & 13.50 & 24.11 & 50.05 \\
         & RMSE & 49.25 & 50.82 & 58.42 & 59.94 & 58.19 & 61.37 & 47.47 & 48.89 & 3.99 & 4.14 & 11.41 & 11.68 & 4.71 & 5.70 & 10.46 & 11.37 & 0.82 & 0.81 & 11.03 & 11.79 & 41.34 & 40.83 & 71.15 & 69.42 \\
         & MAPE & 23.86 & 27.95 & 22.57 & 26.28 & 15.21 & 17.91 & 16.83 & 19.50 & 35.19 & 43.62 & 132.37 & 143.16 & 5.23 & 6.56 & 13.51 & 16.39 & 7.21 & 7.41 & 73.36 & 85.08 & 53.82 & 60.91 & 25.17 & 30.96 \\
        \midrule
        \multirow{3}{*}{FactoST} & MAE & 25.84 & 40.31 & 33.07 & 49.09 & 38.06 & 54.66 & 27.60 & 40.82 & 1.33 & 1.97 & 6.62 & 7.87 & 2.21 & 3.35 & 5.19 & 7.12 & \underline{0.40} & 0.58 & 3.77 & 5.89 & 8.84 & 13.98 & 17.26 & 26.28 \\
         & RMSE & 40.26 & 60.37 & 48.33 & 69.44 & 56.47 & 78.46 & 40.22 & 58.58 & 2.89 & 3.83 & 10.95 & 12.64 & 4.96 & 7.26 & 10.70 & 14.26 & 0.70 & 0.97 & 7.11 & 11.21 & 21.68 & 33.21 & 35.23 & 55.77 \\
         & MAPE & 24.91 & 41.07 & 23.03 & 38.10 & 17.85 & 26.76 & 17.02 & 26.89 & 22.71 & 39.55 & 89.77 & 117.38 & 4.88 & 7.71 & 12.88 & 17.18 & 5.84 & 9.30 & 54.36 & 75.63 & 42.33 & 54.49 & 20.61 & 31.31 \\
        \midrule
        \multirow{3}{*}{UrbanFM} & MAE & \underline{15.62} & \underline{17.03} & \underline{20.95} & \underline{23.00} & \underline{23.71} & \underline{26.39} & \underline{17.70} & \underline{19.19} & \underline{0.71} & \underline{1.00} & \underline{6.06} & \underline{6.05} & \underline{1.68} & \underline{2.11} & \underline{4.23} & \underline{5.06} & 0.44 & 0.49 & \underline{2.12} & \underline{2.81} & 5.60 & 7.74 & \underline{9.57} & \underline{12.92} \\
         & RMSE & \underline{23.24} & \underline{26.30} & \underline{32.21} & \underline{35.26} & \underline{34.99} & \underline{38.93} & \underline{25.80} & \underline{28.57} & \textbf{1.66} & \textbf{2.13} & \underline{10.04} & \underline{10.05} & \underline{3.27} & \underline{4.19} & \textbf{7.27} & \textbf{8.56} & \underline{0.61} & \underline{0.71} & \underline{3.64} & \underline{5.13} & \underline{11.65} & \underline{16.33} & \underline{16.69} & \underline{23.54} \\
         & MAPE & \underline{15.24} & \underline{15.96} & \underline{13.42} & \underline{15.03} & \underline{11.26} & \underline{12.09} & \underline{13.18} & \underline{13.77} & \underline{10.45} & \underline{14.70} & \underline{48.99} & \underline{48.33} & \underline{3.46} & \underline{4.52} & \underline{9.68} & \underline{11.77} & \underline{5.39} & \underline{6.23} & \underline{24.39} & \underline{29.53} & \underline{29.55} & 40.77 & \underline{15.92} & \underline{20.62} \\
        \midrule
        \multirow{3}{*}{\textbf{TopoST}} & MAE & \textbf{13.73} & \textbf{15.29} & \textbf{19.16} & \textbf{21.26} & \textbf{20.11} & \textbf{22.56} & \textbf{15.11} & \textbf{16.79} & \textbf{0.69} & \textbf{0.97} & \textbf{5.55} & \textbf{5.71} & \textbf{1.39} & \textbf{1.74} & \textbf{3.56} & \textbf{4.38} & \textbf{0.27} & \textbf{0.36} & \textbf{1.63} & \textbf{2.12} & \textbf{3.00} & \textbf{4.27} & \textbf{5.54} & \textbf{9.65} \\
         & RMSE & \textbf{21.09} & \textbf{23.78} & \textbf{30.38} & \textbf{33.59} & \textbf{31.25} & \textbf{35.06} & \textbf{23.59} & \textbf{26.50} & \underline{1.74} & \underline{2.17} & \textbf{9.38} & \textbf{9.62} & \textbf{2.83} & \textbf{3.63} & \underline{7.58} & \underline{9.14} & \textbf{0.44} & \textbf{0.59} & \textbf{2.97} & \textbf{3.97} & \textbf{7.75} & \textbf{11.59} & \textbf{11.03} & \textbf{19.51} \\
         & MAPE & \textbf{12.71} & \textbf{14.04} & \textbf{12.40} & \textbf{13.80} & \textbf{8.54} & \textbf{9.67} & \textbf{9.27} & \textbf{10.40} & \textbf{10.01} & \textbf{13.62} & \textbf{46.57} & \textbf{48.00} & \textbf{2.73} & \textbf{3.61} & \textbf{8.39} & \textbf{10.24} & \textbf{3.47} & \textbf{4.70} & \textbf{23.19} & \textbf{28.00} & \textbf{21.59} & \textbf{27.49} & \textbf{9.65} & \textbf{14.98} \\
        \bottomrule
    \end{tabular}%
    }
    \endgroup
\end{table*}

\section{Experiments}
\label{sec:experiments}

\subsection{Experimental Setup}
\label{sec:experimental_setup}

\subsubsection{Datasets and Protocol}
\label{sec:setup_data}
Our primary experiments use the OpenCity benchmark~\cite{li2025open} and follow its evaluation protocol, using 20 source datasets for pretraining and six target datasets for zero-shot evaluation. The datasets cover traffic flow, traffic speed, taxi demand, and bike demand.

\textbf{Pretraining data.} The source collection comprises 13 traffic-flow datasets (CAD4-1 to CAD4-4, CAD7-1 to CAD7-3, CAD8-1, CAD8-2, CAD12-1, CAD12-2, PEMS04, and PEMS08), six traffic-speed datasets (METR-LA, PEMS-BAY, TrafficHZ, TrafficZZ, TrafficCD, and TrafficJN), and one taxi-demand dataset (NYC-TAXI). Together, these datasets contain 10,110 nodes and 151.1 million observations.

\textbf{Target data.} We evaluate three zero-shot scenarios on six datasets excluded from pretraining: \emph{Cross-Region}, using CAD3 and CAD5 (traffic flow) and PEMS07M (traffic speed) to assess transfer to unseen regions in California; \emph{Cross-City}, using TrafficSH (traffic speed) and CHI-TAXI (taxi demand) to assess transfer to Shanghai and Chicago, respectively; and \emph{Cross-Category}, using NYC-BIKE to forecast bike demand, a category absent from pretraining.

\textbf{Protocol.} For pretraining, we use the complete observation sequence within the designated benchmark period of each of the 20 source datasets, without reserving source validation or test sets. We use the final pretraining checkpoint of \textsc{TopoST} for zero-shot evaluation. Each target dataset is split chronologically into training/validation/test sets according to the evaluation protocol: 50\%/10\%/40\% for flow and speed, and 20\%/20\%/60\% for taxi and bike demand. Input--target windows are constructed entirely within individual target splits. Full-shot baselines are trained on the target training sets, with checkpoints selected on the validation sets. Zero-shot models are applied directly to the target test sets without parameter updates. Using the benchmark spatial graphs, zero-shot models predict the next 24 hours from the preceding 24 hours: 288 input steps and 288 output steps for CAD3, CAD5, and PEMS07M at five-minute resolution, or 48 input steps and 48 output steps for TrafficSH, CHI-TAXI, and NYC-BIKE at 30-minute resolution. All methods are evaluated on the same target test splits with the same forecast horizon.

\subsubsection{Baselines}
\label{sec:setup_baselines}
We compare \textsc{TopoST} with three urban foundation models: OpenCity~\cite{li2025open}, UrbanFM~\cite{chen2026urbanfm}, and UniST~\cite{yuan2024unist}. For this OpenCity-benchmark comparison, all four models use the same 20 source datasets and six target datasets. We pretrain UrbanFM and UniST using their official implementations; OpenCity results are taken from its original paper. We also include ten full-shot models: recurrent models TGCN~\cite{Zhao2020TGCN}, AGCRN~\cite{bai2020agcrn}, and MSDR~\cite{liu2022msdr}; convolutional models STGCN~\cite{yu2018stgcn}, GWN~\cite{wu2019graphwavenet}, MTGNN~\cite{wu2020connecting}, and STSGCN~\cite{Song2020spatial}; and attention-based models ASTGNN~\cite{guo2022astgnn}, STWA~\cite{cirstea2022stwa}, and PDFormer~\cite{jiang2023pdformer}. Results for these ten models are also taken from OpenCity~\cite{li2025open}. Table~\ref{tab:zero-shot} distinguishes the two supervision settings.

\subsubsection{Implementation}
\label{sec:setup_hyperparameters}
The following implementation settings apply to the OpenCity experiments.
We use SSE with $d_s=16$ and a 3-layer GIN with hidden and output dimensions of 64. Each GIN layer uses a 2-layer MLP with ReLU and LayerNorm. Dropout with a rate of 0.1 is applied to the outputs of GIN layers 1 and 2. The output of GIN layer 3 is passed through a $64\to8$ linear projection followed by LayerNorm to produce conditioning codes with $d_z=8$. The backbone has $L=2$ blocks with embedding dimension $d=128$, each comprising temporal self-attention, spatial graph convolution, and an FFN. Temporal attention uses 16 heads and dropout of 0.3. The spatial module uses $\alpha=0.05$ and output dropout of 0.1. METIS performs balanced partitioning with a target subgraph size of 16 nodes, yielding subgraphs of 13--18 nodes across the source and target datasets. Non-overlapping temporal patches span 1 hour; patches at 30-minute resolution are zero-padded to 12 slots. We pretrain the model for 5 epochs using Adam with learning rate $10^{-3}$, batch size 16, gradient clipping at 5. Given the cost of multi-dataset pretraining, we report results from a single training run per configuration.

\subsubsection{Evaluation Metrics}
\label{sec:setup_metrics}
We evaluate forecasting performance using 3 widely used metrics: Mean Absolute Error (MAE), Root Mean Square Error (RMSE), and Mean Absolute Percentage Error (MAPE). MAE measures the average absolute difference between predicted and ground-truth values. RMSE is the square root of the mean squared prediction error and gives greater weight to larger errors. MAPE measures the average absolute prediction error as a percentage of the ground-truth value. Lower values of all 3 metrics indicate better predictive performance.

\begin{table}[!t]
    \centering
    \caption{Model sizes and training-cost comparison over the 20 source datasets on a single NVIDIA RTX PRO 6000 Blackwell Workstation Edition GPU (96\,GiB).}
    \label{tab:training_efficiency}
    \begingroup
    \small
    \setlength{\tabcolsep}{3pt}
    \renewcommand{\arraystretch}{1.1}
    \begin{tabular}{@{}lcccc@{}}
        \toprule
        Model & \makecell{Params\\(M)} & \makecell{Batch\\size} & \makecell{Time\\(h/epoch)} & \makecell{Memory\\(GiB)} \\
        \midrule
        UniST               & 3.92  & 16    & 4.50      & 17.51 \\
        OpenCity            & 25.97 & 8     & 5.05      & 94.95 \\
        UrbanFM             & 2.25  & 16    & 17.74     & 12.90 \\
        \textsc{TopoST}     & \textbf{2.16} & 16 & \textbf{3.84} & \textbf{1.28} \\
        \bottomrule
    \end{tabular}
    \endgroup
    \vspace{0.8\baselineskip}
    \caption{Inference-efficiency comparison on CAD5 and CHI-TAXI. Latency is mean $\pm$ standard deviation over 100 complete requests; memory denotes peak GPU usage.}
    \label{tab:efficiency}
    \begingroup
    \small
    \setlength{\tabcolsep}{3pt}
    \renewcommand{\arraystretch}{1.1}
    \begin{tabular}{@{}lcccc@{}}
        \toprule
        \multirow{2}{*}{Model} & \multicolumn{2}{c}{CAD5} & \multicolumn{2}{c}{CHI-TAXI} \\
        \cmidrule(lr){2-3}\cmidrule(l){4-5}
         & \makecell{Latency\\(ms)} & \makecell{Memory\\(GiB)} & \makecell{Latency\\(ms)} & \makecell{Memory\\(GiB)} \\
        \midrule
        UniST & $6.03\pm0.20$ & 0.94 & $5.00\pm0.24$ & 0.90 \\
        OpenCity & $7.59\pm0.10$ & 1.13 & $5.44\pm0.02$ & 1.00 \\
        UrbanFM & $62.22\pm0.21$ & 0.94 & $26.86\pm0.28$ & 0.86 \\
        \textsc{TopoST}-S & $45.43\pm0.36$ & \textbf{0.86} & $16.36\pm0.04$ & \textbf{0.85} \\
        \textsc{TopoST}-P & $\mathbf{3.81\pm1.39}$ & 0.93 & $\mathbf{3.40\pm0.02}$ & 0.87 \\
        \bottomrule
    \end{tabular}
    \endgroup
\end{table}

\subsection{Zero-Shot Forecasting}
\label{sec:zero_shot}

We evaluate zero-shot forecasting by using one day of historical observations to predict the entire following day. At the native sampling resolution of each dataset, the input and forecast windows each contain $T_h=T_f=288$ steps for CAD3, CAD5, and PEMS07M at five-minute intervals, and $T_h=T_f=48$ steps for TrafficSH, CHI-TAXI, and NYC-BIKE at 30-minute intervals. All six target datasets are excluded from pretraining, and the pretrained models forecast directly from target observations and spatial graphs without fine-tuning or parameter updates. Table~\ref{tab:zero-shot} reports results over the full one-day forecast horizon on the target test splits.

As shown in Table~\ref{tab:zero-shot}, \textsc{TopoST} outperforms the existing foundation models OpenCity, UrbanFM, and UniST across all 18 dataset--metric combinations on the 6 target datasets. Averaging across datasets, \textsc{TopoST} reduces MAE, RMSE, and MAPE by 10.81\%, 12.46\%, and 9.16\% compared with OpenCity, and by 9.06\%, 10.11\%, and 9.37\% compared with UrbanFM. These consistent gains with the same source datasets demonstrate stronger zero-shot forecasting performance across heterogeneous urban data. \textbf{The results support the effectiveness of combining shared learning over local subgraphs with topology-conditioned node specialization for prediction on unseen graphs.}

Compared with the 10 full-shot models trained on the target datasets, \textsc{TopoST} achieves the best results in 17 of the 18 dataset--metric combinations, including MAE and RMSE on every dataset, without target-domain training. On TrafficSH, its MAE and RMSE are 26.09\% and 27.52\% lower than those of ASTGNN, the strongest full-shot baseline on these metrics; on CHI-TAXI, the corresponding reductions over STGCN are 40.45\% and 56.08\%. These results demonstrate that \textsc{TopoST} can transfer knowledge learned from heterogeneous source datasets to unseen targets and achieve strong forecasting performance without target-specific training.

\textbf{Validation on the UrbanFM benchmark.}
For supplementary validation, we consider the UrbanFM benchmark~\cite{chen2026urbanfm}, whose official data release provides 80 pretraining datasets and 12 target datasets. We follow its target splits and $12\!\rightarrow\!12$ and $24\!\rightarrow\!24$ forecasting settings. \textsc{TopoST} is pretrained separately for the OpenCity and UrbanFM benchmarks, without transferring checkpoints between them. Table~\ref{tab:urbanfm-zero-shot} compares our results with those reported for UrbanFM, FactoST, and OpenCity in the original UrbanFM paper, and with UniST using its official implementation. \textsc{TopoST} achieves the lowest error in 34 of the 36 dataset--metric combinations under each setting. \textbf{These results demonstrate \textsc{TopoST}'s effectiveness across benchmark protocols and forecasting horizons.}

\begin{table}[!t]
    \centering
    \caption{Component and full-graph ablations. The full \textsc{TopoST} uses SSE. Lower is better; bold marks minima, including ties. MAPE is in percent.}
    \label{tab:ablations}
    \begingroup
    \footnotesize
    \setlength{\tabcolsep}{2pt}
    \renewcommand{\arraystretch}{1.12}
    \resizebox{\columnwidth}{!}{%
    \begin{tabular}{@{}l*{2}{rrr}@{}}
        \toprule
        \multirow{2}{*}{Variant} & \multicolumn{3}{c}{CAD5} & \multicolumn{3}{c}{CHI-TAXI} \\
        \cmidrule(lr){2-4}\cmidrule(lr){5-7}
         & MAE & RMSE & MAPE & MAE & RMSE & MAPE \\
        \midrule
        \textsc{TopoST} (full) & \textbf{9.36} & \textbf{15.82} & \textbf{24.12} & \textbf{1.84} & \textbf{4.19} & \textbf{38.92} \\
        w/o temporal specialization & 9.81 & 16.53 & 25.75 & 1.91 & 4.48 & 40.07 \\
        w/o spatial specialization & 10.28 & 17.28 & 26.20 & 1.94 & 4.55 & 44.00 \\
        w/o node specialization & 9.78 & 16.63 & 25.95 & 1.91 & 4.43 & 41.99 \\
        Topology as features & 9.76 & 16.41 & 25.18 & 1.95 & 4.61 & 39.53 \\
        w/o neighborhood aggregation & 9.98 & 16.86 & 25.35 & 1.88 & 4.39 & 40.35 \\
        w/o adaptive adjacency & 9.69 & 16.54 & 24.60 & 1.91 & 4.47 & 39.75 \\
        \midrule
        Full graph w/ node specialization & 10.34 & 17.67 & 27.28 & 2.08 & 4.98 & 44.48 \\
        Full graph w/o node specialization & 10.58 & 17.90 & 28.56 & 2.19 & 4.90 & 60.87 \\
        \bottomrule
    \end{tabular}%
    }
    \endgroup

    \vspace{0.8\baselineskip}

    \caption{Structural descriptor comparison with other components fixed. The full \textsc{TopoST} uses SSE. Lower is better; bold marks minima, including ties. MAPE is in percent.}
    \label{tab:descriptor_ablations}
    \begingroup
    \small
    \setlength{\tabcolsep}{2pt}
    \renewcommand{\arraystretch}{1.12}
    \begin{tabular*}{\columnwidth}{@{\extracolsep{\fill}}l*{2}{rrr}@{}}
        \toprule
        \multirow{2}{*}{Variant} & \multicolumn{3}{c}{CAD5} & \multicolumn{3}{c}{CHI-TAXI} \\
        \cmidrule(lr){2-4}\cmidrule(lr){5-7}
         & MAE & RMSE & MAPE & MAE & RMSE & MAPE \\
        \midrule
        SSE descriptor & 9.36 & 15.82 & 24.12 & \textbf{1.84} & 4.19 & \textbf{38.92} \\
        RWSE descriptor & 9.34 & 15.83 & 24.20 & 1.85 & 4.21 & 39.79 \\
        HKSE descriptor & \textbf{9.27} & 15.87 & 23.90 & 1.85 & \textbf{4.17} & 39.18 \\
        FRSE descriptor & 9.28 & \textbf{15.76} & \textbf{23.88} & 1.86 & 4.19 & 40.25 \\
        \midrule
        w/o descriptor & 9.56 & 16.18 & 24.69 & 1.88 & 4.27 & 39.49 \\
        \bottomrule
    \end{tabular*}
    \endgroup

    \vspace{0.8\baselineskip}

    \caption{Ablations on partitioning methods and target subgraph sizes (nodes). Lower is better; bold marks column minima. MAPE is in percent.}
    \label{tab:partition_sizes}
    \begingroup
    \footnotesize
    \setlength{\tabcolsep}{2pt}
    \renewcommand{\arraystretch}{1.12}
    \begin{tabular*}{\columnwidth}{@{\extracolsep{\fill}}lc*{6}{r}@{}}
        \toprule
        \multirow{2}{*}{Partition} & \multirow{2}{*}{\makecell{Target\\size}} & \multicolumn{3}{c}{CAD5} & \multicolumn{3}{c}{CHI-TAXI} \\
        \cmidrule(lr){3-5}\cmidrule(lr){6-8}
         & & MAE & RMSE & MAPE & MAE & RMSE & MAPE \\
        \midrule
        \multirow{6}{*}{METIS} & 16 & \textbf{9.36} & \textbf{15.82} & \textbf{24.12} & \textbf{1.84} & \textbf{4.19} & \textbf{38.92} \\
         & 24 & 9.45 & 16.13 & 24.26 & 1.87 & 4.32 & 40.39 \\
         & 32 & 9.50 & 16.08 & 25.92 & 1.91 & 4.43 & 42.32 \\
         & 40 & 9.45 & 16.09 & 24.48 & 2.11 & 5.09 & 46.96 \\
         & 48 & 9.38 & 15.95 & 24.24 & 1.90 & 4.38 & 42.36 \\
         & 56 & 9.50 & 16.31 & 24.66 & 2.01 & 4.84 & 43.59 \\
        \midrule
        KD-tree & 16 & 9.70 & 16.35 & 27.46 & 1.87 & 4.22 & 45.30 \\
        Random & 16 & 10.62 & 18.30 & 27.21 & 1.95 & 4.52 & 43.31 \\
        \bottomrule
    \end{tabular*}
    \endgroup
\end{table}

\subsection{Efficiency and Scalability}
\label{sec:efficiency_scalability}

We examine the efficiency and scalability of \textsc{TopoST} from four perspectives: training efficiency, inference efficiency, inference scaling, and computational analysis. Comparisons with UniST, OpenCity, and UrbanFM focus on model size, runtime, and GPU memory requirements. The computational analysis characterizes how local subgraph processing and node-specific weight generation scale with graph size, connecting the model design to its resource requirements.

\textbf{Training efficiency.}
Table~\ref{tab:training_efficiency} compares model size, training time, and peak GPU memory at the reported batch sizes. \textsc{TopoST} has the fewest parameters (2.16\,M) and the lowest peak memory (1.28\,GiB), using approximately 90\% less memory than UrbanFM, the second-smallest model. Its training time is 3.84\,h per epoch, compared with 4.50\,h for UniST, 5.05\,h for OpenCity and 17.74\,h for UrbanFM. These results show that local subgraph training and topology-conditioned node specialization supports a compact model with low training memory and computation costs.

\textbf{Inference efficiency.}
Table~\ref{tab:efficiency} profiles models with the same architectures as Table~\ref{tab:training_efficiency}. Each request forecasts all nodes from one history window: 211 nodes with $288\!\rightarrow\!288$ steps on CAD5, and 77 nodes with $48\!\rightarrow\!48$ steps on CHI-TAXI. We report two mathematically equivalent execution modes of \textsc{TopoST}: \textsc{TopoST}-S processes subgraphs sequentially, while \textsc{TopoST}-P combines them into a disjoint graph with static graph inputs resident on the GPU. Both preserve the original within-subgraph connectivity, including adaptive adjacency, and include online topology encoding and node-specific weight generation. \textsc{TopoST}-P achieves the lowest latency among the reported implementations, with $11.91\times$ and $4.81\times$ speedups over \textsc{TopoST}-S on CAD5 and CHI-TAXI, while increasing peak memory from 0.86 to 0.93\,GiB and from 0.85 to 0.87\,GiB, respectively. \textsc{TopoST}-S limits graph-computation memory to one subgraph at a time, making it preferable for large graphs under tight memory budgets; \textsc{TopoST}-P trades additional memory for lower latency.

\begin{figure}[!t]
    \centering
    \includegraphics[width=\columnwidth]{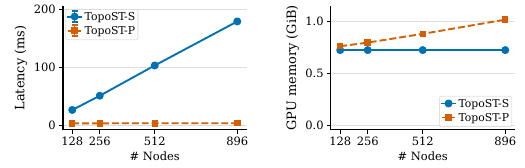}
    \caption{Inference scaling on TrafficSH.}
    \label{fig:efficiency_scalability}
\end{figure}

\textbf{Inference scaling.}
We compare \textsc{TopoST}-S and \textsc{TopoST}-P on TrafficSH graphs formed from node subsets of size $N\in\{128,256,512,896\}$, using identical weights, $48\!\rightarrow\!48$ windows at the native 30-minute resolution, and batch size one. Both modes share the same partitions at each scale, with a target subgraph size of 16. As shown in Fig.~\ref{fig:efficiency_scalability}, \textsc{TopoST}-S's latency grows from 26.52 to 179.56\,ms, while its process memory remains near 0.72\,GiB. \textsc{TopoST}-P takes 3.45--3.67\,ms with 0.76--1.02\,GiB of memory, achieving a $48.93\times$ speedup at 896 nodes. Within this range, parallel execution substantially reduces latency at the cost of a larger memory footprint, while sequential execution favors memory-constrained deployment.

\textbf{Computational analysis.}
Subgraph partitioning reduces spatial computation in training and inference and limits the activations retained for backpropagation. Since the subgraphs form a disjoint cover, $N=\sum_{p=1}^{P}n_p$. With $n_{\max}=\max_p n_p$, we have
\begin{equation}
    \sum_{p=1}^{P} n_p^2
    \leq n_{\max}\sum_{p=1}^{P}n_p
    = n_{\max}N.
    \label{eq:local_spatial_cost}
\end{equation}
For a batch of $B$ temporal windows, evaluating adaptive-graph construction and dense spatial aggregation separately on each subgraph costs $\mathcal{O}((d_z+BT_{\mathrm{p}}d)\sum_p n_p^2)$ per forecasting block, bounded by $\mathcal{O}((d_z+BT_{\mathrm{p}}d)n_{\max}N)$. The corresponding full-graph operations cost $\mathcal{O}((d_z+BT_{\mathrm{p}}d)N^2)$. Backpropagation has the same asymptotic dependence, so partitioning reduces both forward and backward spatial work.

Each training mini-batch contains windows from one subgraph. A block's feature activations and adaptive adjacency therefore require $\mathcal{O}(BT_{\mathrm{p}}n_pd)$ and $\mathcal{O}(n_p^2)$ storage, respectively, compared with $\mathcal{O}(BT_{\mathrm{p}}Nd)$ and $\mathcal{O}(N^2)$ for a full-graph batch. The peak storage for these intermediates is thus controlled by $n_{\max}$, reducing the activation memory needed for training on large source graphs.

Shared parameter pools further control persistent model storage. Across $L$ blocks, the pools contain $\mathcal{O}(Ld_zd^2)$ learned parameters, independent of $N$; their gradients and Adam moment states have the same scaling. The generated node-specific weights require $\mathcal{O}(n_pd^2)$ runtime storage per block but introduce no separate optimizer states.

\subsection{Ablation Studies}
\label{sec:ablation_studies}

We evaluate zero-shot forecasting on CAD5 and CHI-TAXI using models pretrained on the same 20 source datasets.
To test node specialization, we replace temporal node specialization, spatial node specialization, or both with shared operators. Removing spatial specialization also disables adaptive adjacency; \emph{Topology as features} adds topology codes to the shared-operator model's patch embeddings. Two further controls replace GIN with a node-wise MLP or remove only adaptive adjacency, retaining both parameter-generation paths. Full-graph controls train and evaluate on complete city graphs with or without node specialization.

\begin{table*}[!t]
    \centering
    \caption{Forecasting results on CD-DIDI and SZ-DIDI. Lower values are better; bold indicates the best result in each row.}
    \label{tab:didi_few_shot}
    \begingroup
    \footnotesize
    \setlength{\tabcolsep}{2.5pt}
    \renewcommand{\arraystretch}{1.0}
    \setlength{\aboverulesep}{1pt}
    \setlength{\belowrulesep}{1pt}
    \resizebox{\textwidth}{!}{%
    \begin{tabular}{@{}*{14}{w{c}{4.5em}}@{}}
        \toprule
        \multirow{2}{*}{Dataset} & \multirow{2}{*}{Metric $\downarrow$} & \multicolumn{3}{c}{\textbf{Few-shot}} & \multicolumn{3}{c}{\textbf{Zero-shot}} & \multicolumn{6}{c}{\textbf{Full-shot}} \\
        \cmidrule(lr){3-5}\cmidrule(lr){6-8}\cmidrule(lr){9-14}
         & & \textbf{TopoST} & UrbanFM & OpenCity & TopoST & UrbanFM & OpenCity & PDFormer & STWA & AGCRN & MTGNN & GWN & STGCN \\
        \midrule
        \multirow{2}{*}{CD-DIDI} & MAE & \textbf{2.60} & 2.84 & 2.88 & 6.02 & 7.14 & 6.34 & 3.00 & 3.29 & 3.55 & 3.38 & 3.33 & 3.29 \\
         & RMSE & \textbf{3.88} & 4.23 & 4.21 & 7.93 & 9.08 & 9.36 & 4.31 & 4.74 & 5.03 & 4.80 & 4.75 & 4.68 \\
        \midrule
        \multirow{2}{*}{SZ-DIDI} & MAE & \textbf{2.42} & 2.46 & 2.61 & 4.43 & 5.28 & 4.66 & 2.75 & 3.36 & 3.16 & 3.08 & 3.14 & 2.87 \\
         & RMSE & \textbf{3.75} & 3.93 & 4.04 & 6.18 & 7.22 & 7.10 & 4.06 & 4.74 & 4.54 & 4.32 & 4.47 & 4.07 \\
        \bottomrule
    \end{tabular}%
    }
    \endgroup
\end{table*}

The full \textsc{TopoST} achieves the lowest error in all six columns, with no ties at the reported precision (Table~\ref{tab:ablations}). Removing either specialization path or both increases all six errors, and adding topology codes as features does not recover the full model's accuracy. \textbf{Topology-conditioned parameter generation improves zero-shot transfer across both datasets.}
Removing neighborhood aggregation or adaptive adjacency worsens all six metrics, supporting the contribution of both components across the two datasets. Both full-graph controls also underperform the full \textsc{TopoST} on all six metrics, supporting shared learning over local subgraphs.

Table~\ref{tab:descriptor_ablations} compares four 16-dimensional descriptors with an all-one feature control. All four descriptors improve all three CAD5 metrics and CHI-TAXI MAE and RMSE over constant features. SSE serves as the default in our experiments.

\subsection{Partitioning Analysis}
\label{sec:partitioning_analysis}

Table~\ref{tab:partition_sizes} compares METIS target subgraph sizes of 16--56 and three grouping methods at target size 16: METIS, KD-tree, and random grouping. METIS partitions nodes based on graph connectivity, while KD-tree groups them based on distance. Random grouping shuffles nodes and assigns them to groups with the same sizes as METIS-16.
METIS-16 achieves the lowest errors across all six metrics among the configurations in Table~\ref{tab:partition_sizes}, supporting a target subgraph size of 16 within the reported range.
At size 16, METIS outperforms both alternatives on all six metrics. CAD5 MAE is 9.36 with METIS, compared with 10.62 for random grouping and 9.70 for KD-tree grouping; CHI-TAXI MAPE is 38.92\%, compared with 43.31\% and 45.30\%. \textbf{Compact METIS subgraphs preserve the local connectivity used for topology conditioning, thereby supporting cross-dataset learning.}

\subsection{Scaling Behavior Analysis}
\label{sec:transfer_mechanism_analysis}

\begin{figure}[!t]
    \centering
    \includegraphics[width=\columnwidth]{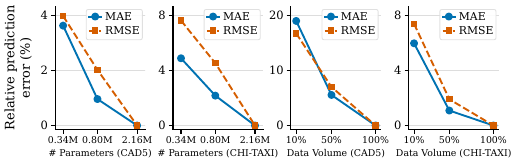}
    \caption{Scaling behavior of \textsc{TopoST}.}
    \label{fig:scaling_behavior}
\end{figure}

We examine parameter and data scalability on CAD5 and CHI-TAXI (Fig.~\ref{fig:scaling_behavior}). \emph{Parameter scalability:} we compare three \textsc{TopoST} variants: Tiny (0.34\,M), Small (0.80\,M), and Main (2.16\,M). Their backbone embedding dimensions are $d=32$, $64$, and $128$, respectively, with the same overall architecture across variants. We plot parameter counts on a logarithmic axis. \emph{Data scalability:} we train the Main variant on chronological prefixes containing 10\%, 50\%, or 100\% of each source dataset's pretraining sequence. To standardize comparisons between MAE and RMSE, the vertical axis reports relative prediction error, $100(E/E_{\mathrm{ref}}-1)$ in percent, with the corresponding Main result in Table~\ref{tab:zero-shot} as $E_{\mathrm{ref}}$. Both metrics decrease as parameter and data scales increase, indicating that greater model capacity and broader source-data coverage improve zero-shot generalization within the evaluated range.

\subsection{Few-Shot Adaptation}
\label{sec:few_shot_adaptation}

We evaluate few-shot adaptation on CD-DIDI and SZ-DIDI using 10\% of each target dataset's training data. As shown in Table~\ref{tab:didi_few_shot}, spatio-temporal foundation models exhibit higher prediction errors in the zero-shot setting than full-shot expert models. Variations in traffic patterns may contribute to this gap. After few-shot adaptation, \textsc{TopoST} substantially reduces its prediction errors, achieving the lowest reported MAE and RMSE on both datasets among the compared models. These results highlight the importance of few-shot adaptation in extending pretrained knowledge to new spatio-temporal data categories with limited target-domain supervision.

\section{Conclusion}
\label{sec:conclusion}

In this paper, we presented \textsc{TopoST}, an urban spatio-temporal foundation model that makes node specialization transferable through a shared topology-to-parameter mapping. Motivated by recurring local structures across datasets and substantial node heterogeneity within subgraphs, \textsc{TopoST} learns over compact local subgraphs and generates node-specific temporal and spatial dependency parameters from identity-independent topological representations. This design enables direct forecasting on unseen graphs without target-specific node embeddings or parameter updates. Experiments on the OpenCity and UrbanFM benchmarks demonstrate strong zero-shot generalization across diverse urban scenarios and forecasting horizons. Ablation and partitioning analyses support the roles of topology-conditioned parameter generation and connectivity-preserving local units, while efficiency results demonstrate low training memory requirements and flexible trade-offs between inference latency and memory. These findings highlight transferable node specialization as an effective approach to sharing predictive knowledge across heterogeneous urban graphs.

\bibliographystyle{IEEEtran}
\bibliography{references}

@inproceedings{liang2025foundation,
  title={Foundation models for spatio-temporal data science: A tutorial and survey},
  author={Liang, Yuxuan and Wen, Haomin and Xia, Yutong and Jin, Ming and Yang, Bin and Salim, Flora and Wen, Qingsong and Pan, Shirui and Cong, Gao},
  booktitle={Proceedings of the 31st ACM SIGKDD Conference on Knowledge Discovery and Data Mining V. 2},
  pages={6063--6073},
  year={2025}
}

@inproceedings{yuan2024unist,
  title={Unist: A prompt-empowered universal model for urban spatio-temporal prediction},
  author={Yuan, Yuan and Ding, Jingtao and Feng, Jie and Jin, Depeng and Li, Yong},
  booktitle={Proceedings of the 30th ACM SIGKDD Conference on Knowledge Discovery and Data Mining},
  pages={4095--4106},
  year={2024}
}

@article{yuan2026uniflow,
  title={UniFlow: a foundation model for unified urban spatio-temporal flow prediction},
  author={Yuan, Yuan and Ding, Jingtao and Han, Chonghua and Sheng, Zhi and Jin, Depeng and Li, Yong},
  journal={IEEE Transactions on Mobile Computing},
  year={2026},
  publisher={IEEE}
}

@article{li2025open,
  title={Open spatio-temporal foundation models for traffic prediction},
  author={Li, Zhonghang and Xia, Long and Shi, Lei and Xu, Yong and Yin, Dawei and Huang, Chao},
  journal={ACM Transactions on Intelligent Systems and Technology},
  year={2025}
}

@inproceedings{chen2026urbanfm,
  title={UrbanFM: Scaling urban spatio-temporal foundation models},
  author={Chen, Wei and Wu, Yuqian and Chen, Junle and Zhou, Xiaofang and Liang, Yuxuan},
  booktitle={Proceedings of the 32nd ACM SIGKDD Conference on Knowledge Discovery and Data Mining V. 2},
  pages={10635--10645},
  year={2026}
}

@inproceedings{zhong2025factost,
  title={Learning to factorize spatio-temporal foundation models},
  author={Zhong, Siru and Qiu, Junjie and Wu, Yangyu and Zou, Xingchen and Rao, Zhongwen and Yang, Bin and Guo, Chenjuan and Xu, Hao and Liang, Yuxuan},
  booktitle={Advances in Neural Information Processing Systems},
  year={2025}
}

@article{zhong2026learning,
  title={Learning to Factorize and Adapt: A Versatile Approach Toward Universal Spatio-Temporal Foundation Models},
  author={Zhong, Siru and Qiu, Junjie and Wu, Yangyu and Liu, Yiqiu and He, Yuanpeng and Rao, Zhongwen and Yang, Bin and Guo, Chenjuan and Xu, Hao and Liang, Yuxuan},
  journal={arXiv preprint arXiv:2601.12083},
  year={2026}
}

@techreport{karypis1997metis,
  author = {Karypis, George and Kumar, Vipin},
  title  = {METIS: A Software Package for Partitioning
                 Unstructured Graphs, Partitioning Meshes,
                 and Computing Fill-Reducing Orderings of
                 Sparse Matrices},
  institution = {University of Minnesota},
  year        = {1997},
  type        = {Technical Report}
}

@inproceedings{shervashidze2009graphlet,
  title={Efficient graphlet kernels for large graph comparison},
  author={Shervashidze, Nino and Vishwanathan, SVN and Petri, Tobias and Mehlhorn, Kurt and Borgwardt, Karsten},
  booktitle={Artificial Intelligence and Statistics},
  pages={488--495},
  year={2009}
}

@article{shervashidze2011weisfeiler,
  title={Weisfeiler-lehman graph kernels},
  author={Shervashidze, Nino and Schweitzer, Pascal and Van Leeuwen, Erik Jan and Mehlhorn, Kurt and Borgwardt, Karsten M},
  journal={Journal of Machine Learning Research},
  year={2011}
}

@article{bai2020agcrn,
  title={Adaptive graph convolutional recurrent network for traffic forecasting},
  author={Bai, Lei and Yao, Lina and Li, Can and Wang, Xianzhi and Wang, Can},
  journal={Advances in Neural Information Processing Systems},
  year={2020}
}

@inproceedings{zhang2017stresnet, 
  title={Deep Spatio-Temporal Residual Networks for Citywide Crowd Flows Prediction}, 
  booktitle={Proceedings of the AAAI Conference on Artificial Intelligence},
  volume={31},
  number={1},  
  author={Zhang, Junbo and Zheng, Yu and Qi, Dekang}, 
  year={2017}
}

@inproceedings{li2018dcrnn,
  title={Diffusion convolutional recurrent neural network: data-driven traffic forecasting},
  author={Li, Yaguang and Yu, Rose and Shahabi, Cyrus and Liu, Yan},
  booktitle={International Conference on Learning Representations},
  year={2018}
}

@inproceedings{yu2018stgcn,
  title={Spatio-temporal graph convolutional networks: A deep learning framework for traffic forecasting},
  author={Yu, Bing and Yin, Haoteng and Zhu, Zhanxing},
  booktitle={International Joint Conference on Artificial Intelligence},
  year={2018}
}

@inproceedings{wu2019graphwavenet,
  title={Graph wavenet for deep spatial-temporal graph modeling},
  author={Wu, Zonghan and Pan, Shirui and Long, Guodong and Jiang, Jing and Zhang, Chengqi},
  booktitle={International Joint Conference on Artificial Intelligence},
  year={2019}
}

@inproceedings{liu2023staeformer,
  title={Spatio-temporal adaptive embedding makes vanilla transformer sota for traffic forecasting},
  author={Liu, Hangchen and Dong, Zheng and Jiang, Renhe and Deng, Jiewen and Deng, Jinliang and Chen, Quanjun and Song, Xuan},
  booktitle={Proceedings of the 32nd ACM International Conference on Information and Knowledge Management},
  pages={4125--4129},
  year={2023}
}

@inproceedings{yuan2025urbandit,
  title={Diffusion Transformers as Open-World Spatiotemporal Foundation Models},
  author={Yuan, Yuan and Han, Chonghua and Ding, Jingtao and Zhang, Guozhen and Jin, Depeng and Li, Yong},
  booktitle={Advances in Neural Information Processing Systems},
  year={2025}
}

@inproceedings{pan2019stmetanet,
  title={Urban traffic prediction from spatio-temporal data using deep meta learning},
  author={Pan, Zheyi and Liang, Yuxuan and Wang, Weifeng and Yu, Yong and Zheng, Yu and Zhang, Junbo},
  booktitle={Proceedings of the 25th ACM SIGKDD International Conference on Knowledge Discovery and Data Mining},
  pages={1720--1730},
  year={2019}
}

@inproceedings{cirstea2022stwa,
  title={Towards Spatio-Temporal Aware Traffic Time Series Forecasting},
  author={Cirstea, Razvan-Gabriel and Yang, Bin and Guo, Chenjuan and Kieu, Tung and Pan, Shirui},
  booktitle={International Conference on Data Engineering},
  year={2022}
}

@inproceedings{lu2022stgfsl,
  title={Spatio-Temporal Graph Few-Shot Learning with Cross-City Knowledge Transfer},
  author={Lu, Bin and Gan, Xiaoying and Zhang, Weinan and Yao, Huaxiu and Fu, Luoyi and Wang, Xinbing},
  booktitle={Proceedings of the 28th ACM SIGKDD Conference on Knowledge Discovery and Data Mining},
  pages={1162--1172},
  year={2022}
}

@inproceedings{yuan2024gpd,
  title={Spatio-Temporal Few-Shot Learning via Diffusive Neural Network Generation},
  author={Yuan, Yuan and Shao, Chenyang and Ding, Jingtao and Jin, Depeng and Li, Yong},
  booktitle={International Conference on Learning Representations},
  year={2024}
}

@inproceedings{fang2025patchstg,
  title={Efficient Large-Scale Traffic Forecasting with Transformers: A Spatial Data Management Perspective},
  author={Fang, Yuchen and Liang, Yuxuan and Hui, Bo and Shao, Zezhi and Deng, Liwei and Liu, Xu and Jiang, Xinke and Zheng, Kai},
  booktitle={Proceedings of the 31st ACM SIGKDD Conference on Knowledge Discovery and Data Mining},
  pages={307--317},
  year={2025}
}

@inproceedings{chiang2019cluster,
  title={Cluster-GCN: An Efficient Algorithm for Training Deep and Large Graph Convolutional Networks},
  author={Chiang, Wei-Lin and Liu, Xuanqing and Si, Si and Li, Yang and Bengio, Samy and Hsieh, Cho-Jui},
  booktitle={Proceedings of the 25th ACM SIGKDD Conference on Knowledge Discovery and Data Mining},
  pages={257--266},
  year={2019}
}

@inproceedings{rampasek2022gps,
  title={Recipe for a general, powerful, scalable graph transformer},
  author={Ramp{\'a}{\v{s}}ek, Ladislav and Galkin, Michael and Dwivedi, Vijay Prakash and Luu, Anh Tuan and Wolf, Guy and Beaini, Dominique},
  booktitle={Advances in Neural Information Processing Systems},
  year={2022}
}

@inproceedings{dwivedi2022lspe,
  title={Graph Neural Networks with Learnable Structural and Positional Representations},
  author={Dwivedi, Vijay Prakash and Luu, Anh Tuan and Laurent, Thomas and Bengio, Yoshua and Bresson, Xavier},
  booktitle={International Conference on Learning Representations},
  year={2022}
}

@article{sun2009hks,
  title={A Concise and Provably Informative Multi-Scale Signature Based on Heat Diffusion},
  author={Sun, Jian and Ovsjanikov, Maks and Guibas, Leonidas},
  journal={Computer Graphics Forum},
  year={2009}
}

@inproceedings{topping2022understanding,
  title={Understanding Over-Squashing and Bottlenecks on Graphs via Curvature},
  author={Topping, Jake and Di Giovanni, Francesco and Chamberlain, Benjamin Paul and Dong, Xiaowen and Bronstein, Michael M.},
  booktitle={International Conference on Learning Representations},
  year={2022}
}

@inproceedings{xu2019gin,
  title={How Powerful are Graph Neural Networks?},
  author={Xu, Keyulu and Hu, Weihua and Leskovec, Jure and Jegelka, Stefanie},
  booktitle={International Conference on Learning Representations},
  year={2019}
}

@inproceedings{hamilton2017graphsage,
  title={Inductive Representation Learning on Large Graphs},
  author={Hamilton, Will and Ying, Zhitao and Leskovec, Jure},
  booktitle={Advances in Neural Information Processing Systems},
  year={2017}
}

@inproceedings{Song2020spatial,
  author = {Song, Chao and Lin, Youfang and Guo, Shengnan and Wan, Huaiyu},
  title = {Spatial-Temporal Synchronous Graph Convolutional Networks: A New Framework for Spatial-Temporal Network Data Forecasting},
  booktitle = {Proceedings of the AAAI Conference on Artificial Intelligence},
  volume={34},
  number={01},
  pages={914--921},
  year = {2020},
}

@ARTICLE{Zhao2020TGCN,
  author={Zhao, Ling and Song, Yujiao and Zhang, Chao and Liu, Yu and Wang, Pu and Lin, Tao and Deng, Min and Li, Haifeng},     
  title={T-GCN: A Temporal Graph Convolutional Network for Traffic Prediction}, 
  journal={Transactions on Intelligent Transportation Systems},  
  year={2020}
}

@inproceedings{wu2020connecting,
  author = {Wu, Zonghan and Pan, Shirui and Long, Guodong and Jiang, Jing and Chang, Xiaojun and Zhang, Chengqi},
  title = {Connecting the Dots: Multivariate Time Series Forecasting with Graph Neural Networks},
  booktitle = {Proceedings of the 26th ACM SIGKDD Conference on Knowledge Discovery and Data Mining},
  pages={753--763},
  year = {2020}
}

@inproceedings{liu2022msdr,
  author = {Liu, Dachuan and Wang, Jin and Shang, Shuo and Han, Peng},
  title = {MSDR: Multi-Step Dependency Relation Networks for Spatial Temporal Forecasting},
  booktitle = {Proceedings of the 28th ACM SIGKDD Conference on Knowledge Discovery and Data Mining},
  pages={1042--1050},
  year = {2022}
}

@inproceedings{jiang2023pdformer,
  title={PDFormer: Propagation Delay-Aware Dynamic Long-Range Transformer for Traffic Flow Prediction}, 
  author={Jiang, Jiawei and Han, Chengkai and Zhao, Wayne Xin and Wang, Jingyuan},
  booktitle={Proceedings of the AAAI Conference on Artificial Intelligence}, 
  volume={37},
  number={4},
  pages={4365--4373},
  year={2023}
}

@article{guo2022astgnn,
  title={Learning dynamics and heterogeneity of spatial-temporal graph data for traffic forecasting},
  author={Guo, Shengnan and Lin, Youfang and Wan, Huaiyu and Li, Xiucheng and Cong, Gao},
  journal={IEEE Transactions on Knowledge and Data Engineering},
  volume={34},
  number={11},
  pages={5415--5428},
  year={2022}
}

@inproceedings{ma2024timeaware,
  title={Learning time-aware graph structures for spatially correlated time series forecasting},
  author={Ma, Minbo and Hu, Jilin and Jensen, Christian S and Teng, Fei and Han, Peng and Xu, Zhiqiang and Li, Tianrui},
  booktitle={2024 IEEE 40th international conference on data engineering (ICDE)},
  pages={4435--4448},
  year={2024}
}

@inproceedings{fang2023stwave,
  title={When spatio-temporal meet wavelets: Disentangled traffic forecasting via efficient spectral graph attention networks},
  author={Fang, Yuchen and Qin, Yanjun and Luo, Haiyong and Zhao, Fang and Xu, Bingbing and Zeng, Liang and Wang, Chenxing},
  booktitle={2023 IEEE 39th international conference on data engineering (ICDE)},
  pages={517--529},
  year={2023}
}

@article{ma2025bist,
  title={Bist: A lightweight and efficient bi-directional model for spatiotemporal prediction},
  author={Ma, Jiaming and Wang, Binwu and Wang, Pengkun and Zhou, Zhengyang and Wang, Xu and Wang, Yang},
  journal={Proceedings of the VLDB Endowment},
  volume={18},
  number={6},
  pages={1663--1676},
  year={2025}
}

@article{han2025scalable,
  title={Scalable Pre-Training of Compact Urban Spatio-Temporal Predictive Models on Large-Scale Multi-Domain Data.},
  author={Han, Jindong and Wang, Hao and Xiong, Hui and Liu, Hao},
  journal={Proc. VLDB Endow.},
  volume={18},
  number={7},
  pages={2149--2158},
  year={2025}
}

@article{lai2023lightcts,
  title={Lightcts: A lightweight framework for correlated time series forecasting},
  author={Lai, Zhichen and Zhang, Dalin and Li, Huan and Jensen, Christian S and Lu, Hua and Zhao, Yan},
  journal={Proceedings of the ACM on Management of Data},
  volume={1},
  number={2},
  pages={1--26},
  year={2023}
}

\end{document}